\documentclass[a4paper,referee,fleqn]{mnras}

\usepackage[T1]{fontenc}
\usepackage{ae,aecompl}

\usepackage{graphicx}	
\usepackage{amsmath}	
\usepackage{amssymb}	
\usepackage{psfig,epsf,rotating,aalongtable,txfonts} 
\numberwithin{equation}{section}

\begin{document}
\title[Statistical analysis of V-types]{Spectral characterization of V-type asteroids - II. \\ A statistical analysis}


\author[S. Ieva et al.]{
S. Ieva,$^{1}$\thanks{E-mail:simone.ieva@oa-roma.inaf.it}
E. Dotto,$^{1}$
D. Lazzaro,$^{2}$
D. Perna,$^{3}$
D. Fulvio,$^{4}$
and M. Fulchignoni$^{3}$
\\
$^{1}$INAF -  Osservatorio Astronomico di Roma, via Frascati 33, I-00040 Monteporzio Catone (Roma), Italy\\
$^{2}$Observatorio Nacional, R. Gen. Jos\'e Cristino, 77 - S\~ao Crist\'ov\~ao, Rio de Janeiro - RJ, 20921-400, Brazil\\
$^{3}$LESIA, Observatoire de Paris, PSL Research University, CNRS, Sorbonne Universit\'es, UPMC Univ. Paris 06, Univ. Paris Diderot, Sorbonne Paris Cit\'e, 5 place Jules Janssen, 92195 Meudon, France\\
$^{4}$Departamento de Fis\'ica, Pontif\'icia Universidade Cat\'olica do Rio de Janeiro, Rua Marques de S\~ao Vicente 225, 22451-900 Rio de Janeiro, Brazil \\
}

\date{Accepted 2015 October 23. Received 2015 October 22; in original form 2015 August 09}

\pubyear{2015}

\label{firstpage}
\pagerange{\pageref{firstpage}--\pageref{lastpage}}
\maketitle

\begin{abstract}
In recent years several small basaltic V-type asteroids have been identified all around the main belt. Most of them are members of the Vesta dynamical family, but an increasingly large number appear to have no link with it. The question that arises is whether all these basaltic objects do indeed come from Vesta. To find the answer to the above questioning, we decided to perform a statistical analysis of the spectroscopic and mineralogical properties of a large sample of V-types, with the objective to highlight  similarities and differences among them, and  shed light on their unique, or not, origin.
The analysis was performed using 190 visible and near-infrared spectra from the literature for 117 V-type asteroids. The asteroids were grouped according to their dynamical properties and their computed spectral parameters compared.  Comparison was also performed with spectral parameters of a sample of HED meteorites and data of the surface of Vesta taken by the VIR instrument on board of the Dawn spacecraft.
Our analysis shows that although most of the V-type asteroids in the inner main belt do have a surface composition compatible with an origin from Vesta, this seem not to be the case for 
 V-types in the middle and outer main belt.    
\end{abstract}




\section{Introduction}

4 Vesta, an asteroid of 530 km diameter orbiting at 2.36 AU, is the biggest ``small body'' to show a basaltic crust. Its basaltic nature was first inferred by McCord, Adams \& Johnson (1970), who found, in the spectrum of Vesta, a deep absorption band near 0.9 $\mu m$, representative of pyroxene composition. Because of the spectroscopic similarity with basaltic achondrites, Vesta was also considered the parent body for Diogenites, Howardites and Eucrites meteorites (Drake, 2001),  collectively known as HED meteorites. With the increase in the number of asteroids discovered in the inner main belt, a dynamical family of 10 members was first identified by Williams (1989).

A major breakthrough  was the discovery that smaller bodies  in orbits between Vesta and the 3:1 mean motion resonance with Jupiter have spectra similar to Vesta (Binzel \& Xu, 1993), confirming a genetic link between them and suggesting a suitable transport mechanism. Another was the discovery from HST images of a giant crater in the south pole of Vesta  (Thomas et al. 1997). The Dawn mission recently identified not just one, but two craters in the south pole of Vesta (Marchi et al. 2012), that with all probability are the origin of Vesta's dynamical family. The scenario was then complete: several impacts created a  swarm of basaltic fragments, some forming the dynamical family while others, due to collisions and dissipative effects, were injected into strong resonances, who pumped their eccentricities. Some were ejected from the Solar System or fell directly into the sun (Farinella et al. 1994). Others, extracted from a close encounter with a terrestrial planet, became Near-Earth Asteroids (NEAs) (Cruikshank et al. 1991) or Mars-crossers (Ribeiro et al. 2014). Finally, some of them ended up colliding on Earth and were recovered as HED meteorites.

Previous dynamical and observational works confirmed the above scenario characterized by the diverse steps of the evolution of fragments from Vesta. However, the discovery of a basaltic asteroid in the outer main belt (Lazzaro et al. 2000) can be regarded as the iceberg tip on the presence and extension of differentiated material in the asteroid belt. Nowadays basaltic asteroids with no dynamical link with Vesta have been discovered all over the main belt (Nesvorn\'y et al. 2008): some of them can be traced back to Vesta trough dynamical pathways that involve nonlinear secular resonances  (Carruba et al. 2005); others could derive from a distinct differentiated body than Vesta.
Moreover laboratory studies on the oxygen isotopic composition of HED meteorites suggest that not all can be related to a unique parent body (Scott et al. 2009).  

Asteroids have been classified as V-type if showing a spectrum similar to that of Vesta (Tholen \& Barucci 1989, Bus \& Binzel 2002, DeMeo et al. 2009) or ``putative'' V-type if presenting compatible photometric colors (Roig \& Gil-Hutton 2006, Carvano et al. 2010, DeMeo \& Carry, 2013). Extensive numerical simulations  of the dynamical evolution of Vesta's ejected fragments over timescales comparable to the family age have shown that a relatively large fraction of the original Vesta family members may have evolved out the family borders, and be considered as ``fugitives'' (Nesvorn\'y et al. 2008). However, from a dynamical point of view it is quite difficult to explain large V-type objects in the middle and outer main belt.  According to Roig et al. (2008) the probability of an asteroid with a mean diameter larger than 5 km to evolve from the Vesta family, cross the 3:1 mean motion resonance with Jupiter, and reach a stable orbit in the middle belt is almost 1\%. It is noteworthy that the estimated diameters (assuming an albedo of 0.4 typical for basaltic surfaces) for the middle belt asteroids 10537 1991 RY16 and 21238 Panarea are about 7 and 5 km, respectively, while for the outer belt asteroid 1459 Magnya, thermal observation have determined a diameter of 17 km (Delbo et al. 2006).  Moreover, these are the three unique V-type asteroids in the middle and outer belt whose basaltic nature has been confirmed through visible and near-infrared spectra. 

In the last decade several works have been devoted to the observation of V-type asteroids, in order to answer the question of their origin (Duffard et al. 2004; Moskovitz et al. 2008a, 2008b, Burbine et al. 2009, Moskovitz et al. 2010, De Sanctis et al. 2011a, 2011b).  
In most of these works the focus has been to identify mineralogical differences and/or similarities between V-type asteroids belonging and not belonging to the Vesta dynamical family. No definitive conclusion has been reached by the diverse authors, with the only exception of 1459 Magnya, which mineralogy has been shown to be distinct from that of  Vesta (Hardersen, Gaffey \& Abell 2004).
For all the basaltic material in the main belt of asteroids with no dynamical connection with  Vesta it is possible to consider three plausible scenarios (Moskovitz et al. 2010):

\begin{itemize}
\item These objects were removed at the epoch of the Vesta family formation and migrated to their current position via some still unknown dynamical mechanism: in this case they should not appear spectroscopically different from Vesta family asteroids. 
\item These objects were removed from 4 Vesta in the early phases of the Solar System formation and were scattered to their current location due to the sweeping of mean motion resonances in the region: in this case they would represent an older population of Vesta family objects and might be spectroscopically distinct.
\item These objects are fragments of the basaltic crust of other differentiated bodies which were disrupted in the early phases of the asteroid belt evolution: in this case they would likely be spectroscopically and mineralogically distinct.  
\end{itemize}

In order to rule out which one of the above hypothesis is not supported by the observations we performed a statistical analysis of the spectroscopic and mineralogical properties of the whole sample of V-type spectra available in literature. To highlight   similarities and differences among the objects included in this sample of V-types, the largest one ever collected and analysed, and shed light on their possible Vestan origin, we computed and analysed several spectral parameters in the visible and/or near-infrared ranges.
Asteroids were grouped according to their dynamical properties and their computed spectral parameters were compared with each other, with those of a sample of HED meteorites and with spectral parameters of the surface of Vesta as taken by VIR instrument on board the Dawn spacecraft.
In the next section we describe the selected sample, while the statistical analysis of the data is given in section 3.  The study of the mineralogy for the sub-sample with infrared data is given in section 4. Finally, we conclude discussing the results we obtained and their implications on the study of V-type asteroids in the main belt.

\section{The V-type sample}
The selected sample is composed of 190 spectra in the visible and/or near-infrared range for 117 asteroids classified as V-type according to the most recent taxonomy (DeMeo et al. 2009).
In order to better investigate similarities between different populations we divided our sample in six groups according to their dynamical properties. The highest concentration of V-type asteroids is found in the inner main belt (at semimajor axis \emph{a}$<2.5$ AU), where we identified four dynamical groups: vestoids or Vesta family objects, fugitives, low-inclination (low-i) and inner other (IOs). Outside the inner main belt two other V-type groups were defined: Near-Earth Asteroids (NEAs) and Middle/Outer V-types (MOVs, at \emph{a}$>2.5$ AU). Objects outside  the limits of the Vesta dynamical family (fugitives, low-i, IOs and MOVs) are often grouped together in  literature as non-vestoids. Therefore we considered:

\begin{itemize}
\item A \textbf{vestoid} is a V-type member of the Vesta dynamical family, as defined  by Nesvorn\'y\footnote{http://sbn.psi.edu/pds/resource/nesvornyfam.html} using the Hierarchical Clustering Method (HCM). 
\item A \textbf{fugitive},  following the definition of Nesvorn\'y et al. (2008), is a V-type asteroid with \emph{a} $<$  2.3 AU and comparable \textit{e} and \textit{i} with the Vesta family. 
\item Also according  to Nesvorn\'y et al. (2008) a low-inclination (\textbf{low-i}) is a  V-type asteroid having \emph{i} $< 6^{\circ}$ and $2.3<\emph{a}<2.5$ AU.
\item The remaining V-type asteroids in the inner  main belt were identified as \textbf{IOs.}
\item A \textbf{NEA} is a  V-type asteroid  in the near-Earth region (with a perihelium q $<$ 1.3 AU).
\item A \textbf{MOV} is a V-type asteroid in the middle and outer main belt ($\emph{a}>2.5$ AU).
\end{itemize}

Our final  sample is composed of 44 vestoids, 15 fugitives, 23 low-i, 9 IOs, 4 MOVs and 22  NEAs. 
 In Tab. \ref{obs} the orbital parameters (\textit{a}, \textit{e} and \textit{i}) are given for each asteroid of the selected sample, as well as the solar phase at the  time of the observation in the visible and/or in NIR, the reference from which  the spectra were taken and the assigned dynamical group. 
 Multiple spectral observations were used to estimate the influence of observational systematic errors. The independent analysis on the spectra of the same object taken with different telescopes and different observational conditions has excluded the possibility of systematic errors.

Since the principal aim of this analysis is to check if the different V-types  across the main belt and near-Earth region share the same properties as those objects  that very plausibly come from Vesta (i.e the vestoids), we considered a \textbf{control sample} defined by those vestoids which have both visible and near-infrared spectra obtained at solar phase angles which differs between 1$^{\circ}$ and 12$^{\circ}$.  The last condition  is important to guarantee that the combination of visible and near-infrared spectrum is not affected by observational and geometrical effects.  
For some vestoids although (2468, 4815, 6159, 8693, 10037, 42947, 66268) due to the large scattering in the data some parameters were not computed, therefore they were excluded from the control sample. The final control sample is  marked with an asterisk in Tab. \ref{obs}.

\section{Statistical analysis}
\subsection{Visible range}

We performed the statistical analysis on visible spectra for our sample of 85 V-types focusing on three parameters: the reflectivity gradient in the 5000 - 7500 \textup{\AA} and 8000 - 9200 \textup{\AA} range (\emph{slopeA} and \emph{slopeB} respectively), and the \emph{apparent depth} (the ratio between the reflectivity at 7500 and 9000 \textup{\AA}). These parameters are chosen since they could characterize the position and the shape of the 0.9 $\mu m$ absorption band for V-type asteroids having visible spectra. A steeper slopeA could be indicative of weathered surfaces (Fulvio et al. 2012, 2015),  while a deeper apparent depth could be indicative of bigger grain size (Cloutis et al. 2013) or of a fresh unweathered pyroxene.
The computed spectral parameters for each asteroid with a visible spectrum are given in Tab. \ref{vistab}, together with the mean value for each dynamical group and the control sample. For some asteroids,  due to the large scattering in the data, some parameters could not be computed.

\begin{figure}
   \centering
   \includegraphics[angle=0,width=17cm]{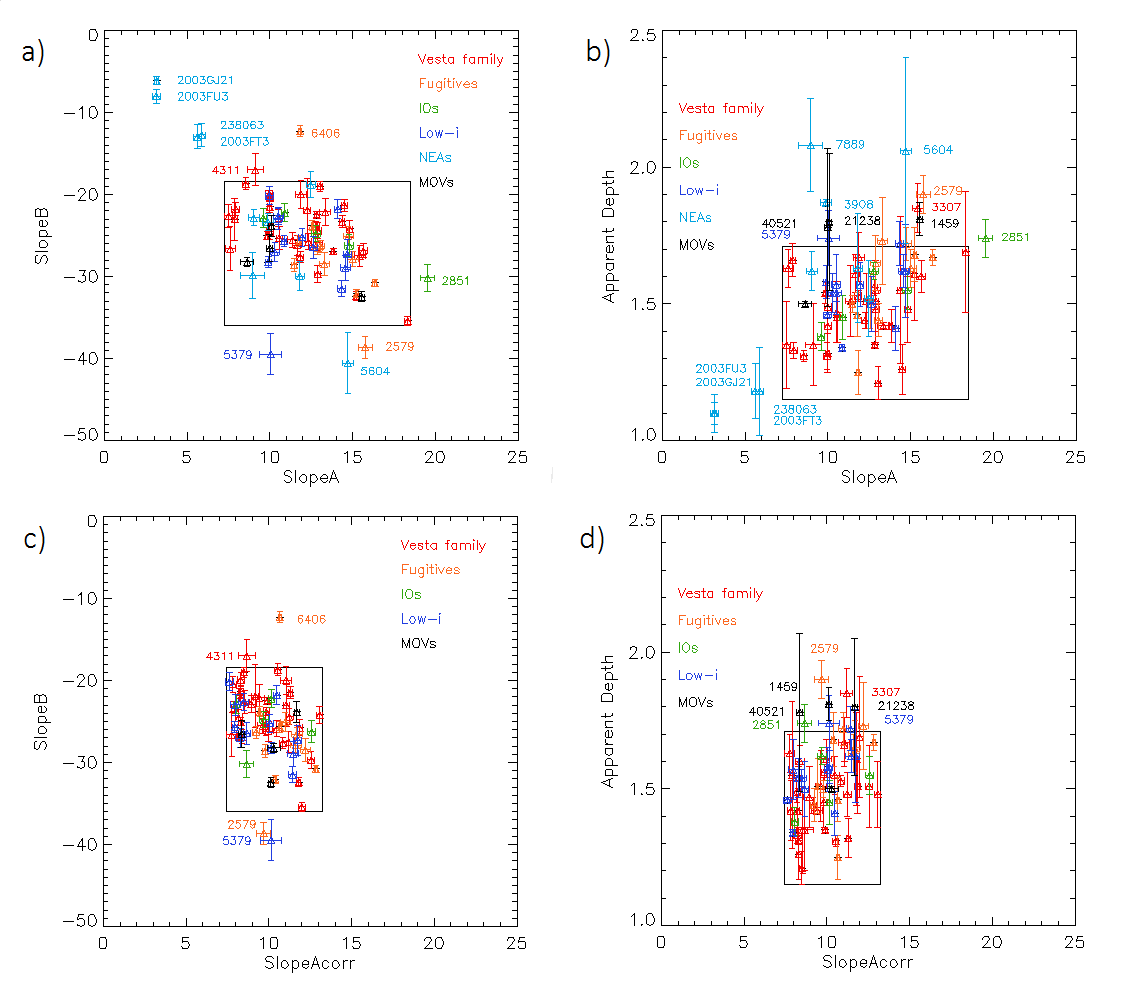}
      \caption{ a-b) SlopeA, SlopeB and Apparent Depth for the whole sample of V-type asteroids with visible spectra  divided by dynamical groups: Vestoids, Fugitives, IOs, Low-i, NEAs and MOVs. c-d) SlopeA after the the empirical phase angle correction applied with the Reddy et al. (2012) relation, SlopeB and Apparent Depth for Vestoids, Fugitives, IOs, Low-i and MOVs. No correction was applied to NEAs, observed at phase angles $> 30^{\circ}$.
      The box is defined by the values of the spectral parameters belonging to the control sample: V-types observed in the visible and NIR range under similar observational conditions and belonging to the Vesta dynamical family. Outsiders are discussed in the text.
      }
         \label{visdinamica}
   \end{figure}

In Fig. \ref{visdinamica}a - \ref{visdinamica}b the whole sample of 85 V-types analyzed in the visible range is grouped according to their dynamical properties. The control sample above defined, taking into account the errors bars,  marks a compact region: slopeA can vary between 7.26 and 18.51\%$/10^3$ \textup{\AA}, with an average value of 12.12\%$/10^3$ \textup{\AA}; slopeB is found in a range between -18.41 and -35.98\%$/10^3$ \textup{\AA}, with an average of -24.60\%$/10^3$ \textup{\AA}; the average apparent depth is 1.46, and it varies between 1.15 and 1.72. 
In Fig. \ref{visdinamica}a we plotted slopeA vs slopeB. The majority of V-types cluster inside the region defined by the control sample and there is no apparent discontinuity between the spectral parameters computed for vestoids, fugitives, low-i and  IOs. 
NEAs show the lower slopeA and the greatest and lowest slopeB.
In Fig. \ref{visdinamica}b we report slopeA vs apparent depth. Most of our database of V-types cluster in the region defined by the control sample, while some asteroids, mostly NEAs and MOV objects, have higher apparent depth than the average. The extremely low slopeA found for 4 objects (238063, 2003FT3, 2003FU3 and 2003GJ21), could be indicative of unweathered  surfaces or freshly regardened material, although in this region these spectra are particularly noisy. 
  
It should be noticed however that, due to a phase angle effect,  some objects could appear redder than they actually are.
To correct for phase angle effect we applied to main belt V-types (observed at $\alpha < 30^{\circ}$, see Tab. \ref{obs}) a relation found by Reddy et al. (2012) for Vesta, which in principle can be used for asteroids who share the same mineralogy (e.g. the V-type asteroids). It is important to stress out that this relation was originally found for Vesta, and this correction applied to our sample is more qualitative than quantitative.
For NEAs, observed between $39^{\circ} < \alpha < 93^{\circ}$, we do not apply any correction, since  the Reddy et al. (2012) empiric relation works well within phase angles $\alpha < 30^{\circ}$, and we did not find any evident correlation between slopeA and phase angle.
In Fig. \ref{visdinamica}c -  \ref{visdinamica}d the whole sample of main belt V-types is plotted after the empirical correction for the phase angle effect. The greater slopeA shown by some objects could be indicative of weathered surfaces. 
The corrected control sample has now a slopeA between 7.45 and 13.25\%$/10^3$ \textup{\AA}. Few objects are plotted above the control sample, with a lower slopeB in Fig.  \ref{visdinamica}c.  Due to the absence of infrared counterpart 4311 could  belong to a different taxonomic group. In fact, outside the region defined by the control sample there is also 6406, which is reported in literature as V-type, but from a careful analysis of its NIR spectrum it show shallower band depths, and could be classified as Sv-type.  Objects below the region defined by the control sample show a greater slopeB, with possible unweathered surfaces; however it is noteworthy to say that from the comparison of its NIR spectrum 5379 also seems to belong to the S-complex.
In Fig.  \ref{visdinamica}d for two MOV objects (21238 and 40521) the experimental errors in the depth determination, due to the low S/N in this region,  are too big to exclude a compatibility with the control sample zone. Other V-types outside the control sample (1459, 2579 and 3307) show an apparent depth greater than 1.80, which in principle could be due to a bigger grain size, fresh pyroxene  or a different mineralogy.  In particular 1459 is a MOV object and its position in the outer main belt could point to a different parent body and mineralogy than Vesta.

\begin{figure}
   \centering
 \includegraphics[angle=0,width=17cm]{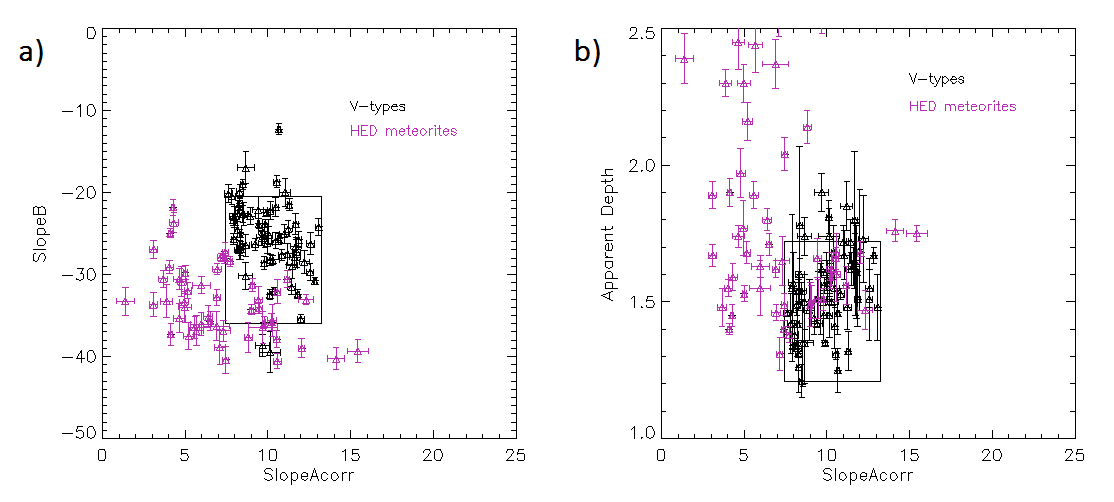}
      \caption{SlopeA \textbf{vs} SlopeB and SlopeA \textbf{vs} Depth for a sample of HED meteorites and V-type objects corrected for the phase angle effect.
      Note that, even after the correction, these plots show that V-types are intrinsically redder then HED suite.}
         \label{visreddy}
   \end{figure}

In order to verify if V-types have experienced space weathering we compared the whole sample of V-types corrected for the phase angle effect with a sample of HED meteorites taken from the RELAB database\footnote{http://www.planetary.brown.edu/relabdata/} and reported in Tab. \ref{vished}.
V-type asteroids have shallower apparent depth than HED (Fig.  \ref{visreddy}), which might be attributed to space weathering. Moreover,  the majority of the asteroids have SlopeA greater than the majority of the HED, thus suggesting that V-types are intrinsically redder, having experienced a certain degree of space weathering alteration.

\subsection{Near Infrared range}
V-type asteroids are characterized in the NIR range by the presence of two deep absorption features, due to pyroxene minerals, at 0.9 and 1.9 $\mu m$, hereafter BI and BII.  These bands are caused by the $Fe^{2+}$ electronic transitions in the M1 and M2 crystallographic sites of pyroxene structure (Burns 1993). 
According to laboratory experiments, two parameters are the most diagnostic to infer compositional properties: {\bf band minimum} and {\bf band separation} (BII minimum - BI minimum). Cloutis et al. (1990) discovered  that both BII minimum and band separation increase with the increasing iron content.
For 71 V-types of our sample  having only NIR spectra we computed BI and BII spectral parameters. For 11 objects, due to large scattering in the data in the 1-2 $\mu m$ region, it was not possible to compute BI and BII minimum. 
 We computed band minima using standard procedures, fitting each band with $2^{nd}$ order polynomial fit. Errors were computed using a Monte Carlo simulation, randomly sampling data 100 times and taking the standard deviation as incertitude. BI minima, BII minima and band separations are shown in Tab. \ref{nirtab}.

 \begin{figure}
   \centering
   \includegraphics[angle=0,width=17cm]{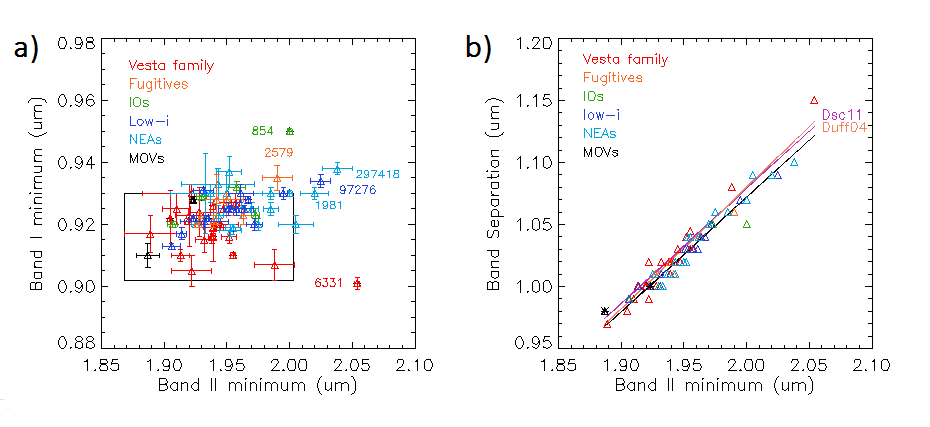}
      \caption{Analysis of near-infrared spectra of V-type asteroids. a) BII vs BI minimum for different dynamical groups (Vestoids, Fugitives, IOs, Low-i, NEAs and MOVs); the box delimits the spectral parameters for the control sample, defined as above. b) BII minimum vs Band separation, divided by dynamical groups. The black line represents our fit, computed from a linear regression model. We also report the linear relation found by Duffard et al. (2004) and De Sanctis et al. (2011b).}
         \label{nir}
   \end{figure}

In Fig. \ref{nir}a we plot BII minima vs BI minima for all of the NIR database.  Vesta family members seem to regroup at shorter BI and BII minima, inside the region defined by the control sample, although 6331 shows higher BII minimum: this is probably due to the low S/N noise in the BII region for this spectrum. 
 Fugitives and IOs  are in good agreement with Vesta family members, although two objects (854, 2579) shows unusual higher BI/BII minima.  Low-i asteroids seem to be compatible with the control sample region, although their distribution is slightly shifted towards longer BI/BII minima wavelenghts. MOVs and NEAs seem to have a different behavior, with one  of the three MOV confirmed in the NIR range (21238) having the lowest BI/BII minima,  while NEAs regroup at longer BI/BII minima respect to Vesta family.
This result is also confirmed computing the weighted mean and the median absolute deviation (MAD) for each dynamical class, reported at the end of Tab. \ref{nirtab}. There seems to be no difference among the BI minima values computed for the different classes, although when comparing BII minima and Band separation values NEAs and  MOVs in particular show unusual values that differ from other classes.

In Fig. \ref{nir}b we plotted BII minima versus band separations.  Data show a clear linear trend, given by:

\begin{equation}
 y = 0.926395 x - 0.780845
\end{equation}

\noindent this correlation, based on the largest dataset of V-types ever collected, confirms what found by Cloutis et al. (1990) on a sample of meteorite spectra, and it is an improvement of the linear fit found by  De Sanctis et al. (2011a), Duffard et al. (2004) and De Sanctis et al. (2011b),  for 18, 14 and 12 V-type asteroids respectively. 
Different dynamical groups seem to range overall the general trend. Most of the Vesta family members appear to have lower band separation, 
showing low amount of iron. V-type NEAs seem to range all over the linear trend, showing both low and high iron content: this could be in agreement with a balance between  space weathering (Pieters et al. 2000) and regardening processes (Binzel et al. 2010) experienced by NEAs.
The linear fits computed for each dynamical class all seem to agree to the general relation found for the global sample of V-types adding the error bars, computed in this case with a maximum/minimum slope method.

\section{Mineralogy}
To further investigate possible differences between different dynamical groups of V-types we conducted a mineralogical analysis using different tools. First we used the Modified Gaussian Model (Sunshine, Pieters \& Pratt 1990) to compare the relative percentage of orthopyroxene and clinopyroxene  in basaltic assemblages. Then we performed a comparison with meteorites in order to characterize the predominant lithology (diogenites, howardites or eucrites) among the analyzed V-types. We used the band depth as a proxy to infer some constraints on the grain size of the surface of V-type asteroids. Finally, we computed the molar composition of iron and calcium content starting from the position of BI and BII center.

\subsection{Modified Gaussian Model}
Basaltic material appears in nature as a single component or as a mixture of two kinds of pyroxene (orthopyroxene, OPX and clinopyroxene, CPX). 
The two models (OPX and OPX+CPX) can be discerned throughout a careful spectral analysis, usually performed with the Modified Gaussian Model (Sunshine et al. 1990). This basic approach deconvolves absorption features into discrete mathematical distributions (modified gaussians), each described by a centre, width and strenght. While band centres and widths do not change, the relative strength of the two major absorption bands of OPX and CPX near 1 and 2 $\mu m$ vary systematically with abundance. The Component Band Strength Ratio (CBSR), defined as the ratio of the strength of the major bands of orthopyroxene and clinopyroxene, vary logarithmically with the OPX/CPX  composition (Sunshine \& Pieters 1993), therefore it can be used to infer mineralogy properties.

We started by deconvolving the spectra of  38 V-types of the  sample, with both visible and NIR spectra and  high S/N in the 1 and 2 $\mu m$ region, using only orthopyroxene input parameters (six absorption band centers, widths and strengths). It is important to start the initial fit with the minimum number of bands, because the fit will always improve with a greater number of bands. The MGM outputs a wavelength dependent RMS error, which show characteristic features diagnostic for missing bands (e.g. the presence of an offset is typical for poorly modeled fits, for further details see Fig. 8 from Sunshine \& Pieters 1993). If a peak error offset from band centers was present we considered the fit not acceptable and
 we restarted the deconvolution using a mix of 75/25 OPX/CPX input parameters (eight absorption centres, widths and strenghts) derived in Sunshine \& Pieters (1993), and following the same procedure. Then we computed the CBSR for the 1 and 2 $\mu m$ region, and related to the CPX/OPX percentage. For six V-types (1933, 3268, 3498, 3968, 4038 and 11699),  due to the large scattering of the data and the possible presence of additional bands (i.e. plagiocase, olivine... etc.) 
 we didn't find an acceptable fit with neither one of the two models, and they will be excluded from the following analysis. The obtained results are summarized in Tab. \ref{mgm1}.

15 V-types were best modeled using only a single pyroxene model (Tab. \ref{mgm1}.1) and they appear to have a composition of orthopyroxene. 5111 Jacliff was modeled using an additional band to account for the M1 site absorption (Burns 1993), often masked by the most prominent M2 site absorption.   
Other 17 V-types of our sample were modeled using both OPX and CPX absorption bands (Tab. \ref{mgm1}.2), which relative strengths therefore can be used to estimate the proportion of clinopyroxene in the sample.
There is no homogeneity in the V-type sample, with relative percentage CPX/(OPX+CPX) ranging from 20\% (809 Lundia) to almost 68\% (854 Frostia and 1981 Midas). The two objects belonging to the Vesta family seem to have a similar  composition, with a 38-45\% of CPX; on the contrary non-vestoid objects show a wide spread of CPX percentage (from 20 to 68\%) with no evident correlation with the dynamical group. The average values computed for Vesta family and non-vestoid objects are however in agreement, showing a 39$\pm$8\% and 41$\pm$14\% of CPX respectively. NEAs show instead a distribution of CPX percentage slightly shifted toward higher values, with two NEAs (1981 and 6611) having a a great CPX composition (61-68\%) and an average CPX percentage of 56$\pm$10\%.
Our results are also in agreement with the recent MGM analysis of Mayne et al. (2011), although for asteroid 2763 we  computed a higher value for CPX percentage. It should be considered however that the estimation of CPX abudance from relative band strenghts are within $\pm$9\% of their actual values (Mayne et al. 2009).

\subsection{HED comparison}
Among V-types with both visible and infrared spectra  we computed band centres and band depths for 38 objects. Other V-types with VNIR spectra have low S/N, generally in the 2 $\mu m$ region, and no parameter was computed.
Band centres  were evaluated 
first removing the continuum between the two local maxima at 0.7 and 1.2-1.4 $\mu m$ and then fitting each band with a $2^{nd}$ order polynomial fit:
BI was fitted in the 0.8 - 1.1  $\mu m$ region, while BII was fitted in the 1.6-2.3 $\mu m$ region. Band depths were computed dividing the reflectance at the band centre by the reflectance of the continuum at the same wavelength, as done in De Sanctis et al. (2012). Spectral parameters were computed on V-type asteroids and on a set of spectra of HED meteorites taken from the RELAB database, and they are shown in Tab. \ref{vnirtab} - \ref{hd}.

In principle, to compare band centre positions for main belt asteroids and meteorites a temperature correction should be applied, because meteorites are analyzed at room temperature, while asteroids have typically lower temperatures, depending on their distance from the sun. However, we decided to not apply any corrections since several recent works (Burbine et al. 2009, Longobardo et al. 2014) has shown that  on pyroxene assemblages the shift in the wavelength position due to temperature are very small.

\begin{figure}
   \centering
   \includegraphics[angle=0,width=17cm]{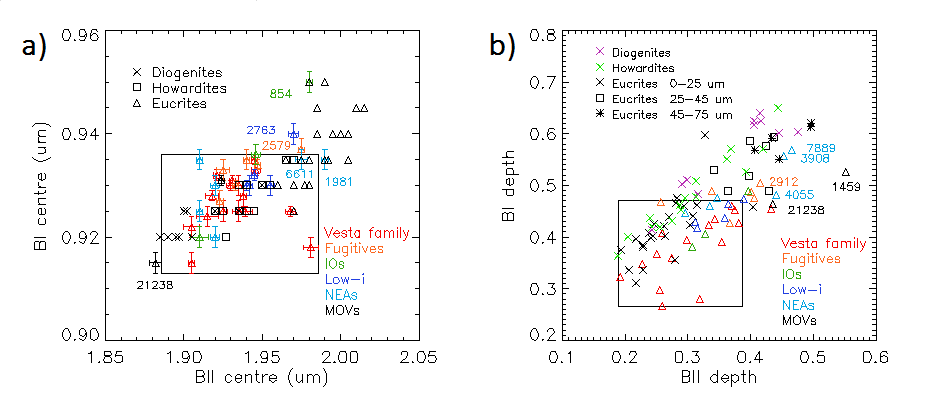}
      \caption{Analysis of VNIR spectra for V-type asteroids. a) BII vs BI centres divided by dynamical groups, along with band parameters for a sample of diogenites, howardites and eucrites. b)  BII vs BI depths for VNIR V-types divided by dynamical groups. We also reported values for eucrites of different grain sizes, taken from the RELAB database. }
         \label{vnir}
   \end{figure}

In Fig. \ref{vnir}a we compared band centres for different dynamical groups, along with the same spectral parameters computed for the HED suite. 
 The dynamical groups seems to range from diogenites to eucrites with no apparent clustering: among five V-types showing a probable eucritic composition, there are two NEAs (1981 and 6611), one IO (854), one fugitive (2579) and one low-i (2763). A few V-types  have spectral parameters compatible with pure diogenite, while the majority of V-types have spectral parameters compatible with howardites. This is in good agreement with the recent findings from the Dawn spacecraft, which orbited Vesta in 2011-2012 and spectroscopically mapped almost the entire surface, computing an average composition for Vesta similar to howardites (De Sanctis et al. 2012). 
 We also confirmed the spectral analogy found with howardites for the fugitive 809 Lundia (Birlan et al. 2014).

The depth of the absorption feature around 1 and 2 $\mu m$ can be used as a proxy to infer regolith properties of the surfaces of V-type asteroids (Hiroi et al. 1995). It is known that pyroxene band depths in HED meteorites increase with the increasing grain size (Cloutis et al. 2013), although laboratory experiments have shown that irradiation processes simulating  space weathering can lower the band depths (Vernazza et al. 2006, Fulvio et al. 2012), while regardening processes can uplift fresh unweathered pyroxene (Binzel et al. 2010).
We compared our database of V-types with eucrites of different grain size taken from the RELAB database. For diogenites and howardites we considered only meteorite samples with grain sizes $< 25 \, \mu $m, since for both of them only one sample of grain sizes $> 25 \, \mu $m is available on RELAB database.  
The majority of our V-type sample seems to be compatible with howardites and eucrites of a grain size $< 25 \, \mu $m (Fig. \ref{vnir}b), although the fugitive 2912 and the NEA 4055 are plotted  in a region compatible with eucrites of a grain size in the 25-45 $\mu $m range, while other NEAs (3908, 7889) show parameters similar to eucrites of grain size $45 -75 \, \mu$m. The two MOVs 1459 and, to some extent 21238, seem to differ to the general trend, which could be symptomatic to a different mineralogy.  
In order to verify if the two populations here considered (the V-type sample and the eucrites $< 25 \, \mu $m) belong to the same class we performed a t-test (Student 1908). 
The difference between the two population is not significant below 1\%, meaning that there is no evidence to suggest that these two populations have different means, and confirming Hiroi et al. (1995) results using a considerable larger dataset. The t-test show also a significant difference between V-types and the $25-45$ and $45-75 \, \mu$m grain size eucrites.

\subsection{Iron-calcium content}
Band centres are the most accurate diagnostic spectral parameters to infer mineralogy in V-type asteroids. Over the last decades several authors studied the relationship between these parameters and olivine and pyroxene mineralogy  (Adams 1974, King \& Ridley 1987; Cloutis \& Gaffey 1991). 
It is known that the position of band centres in orthopyroxene assemblages shifts to longer wavelengths with increasing iron contents, while for clinopyroxene the same phenomenon happens for increasing calcium content (Gaffey et al. 2002).
For this reason, from the position of band centres at 1 and 2 $\mu m$ it is possible to infer the molar contents of calcium (wollastonite, [Wo]) and iron (ferrosilite, [Fs]) of the observed V-types, throughout equations derived on laboratory experiments. 
The Gaffey et al. (2002) equations work well in a wider range of pyroxene samples, with error bars of the order of 4-5\%. The Burbine et al. (2009) equations are empirical relations calibrated in laboratory on  assemblages which reproduce the composition of typical V-types (olivine, orthopyroxene and clinopyroxene). The error bars are in this case in the range of 1-4\%. 

Once computed BI and BII centre with a polynomial fit, following procedures similar to the ones described before,  we applied the Burbine et al. (2009) equations to compute [Wo] and [Fs] contents:

\begin{equation}
Wo \, (\pm 1) = 396.13 \times BI_{cen}  \,  (\mu m) - 360.55 
\end{equation}
\begin{equation}
Wo \, (\pm 1) = 79.905 \times BII_{cen}  \,  (\mu m) -148.3
\end{equation}
\begin{equation}
Fs \, (\pm 3) = 1023.4 \times BI_{cen}  \, (\mu m) - 913.82 
\end{equation}
\begin{equation}
Fs \, (\pm 3) = 205.86 \times BII_{cen}  \,  (\mu m) - 364.3 
\end{equation}
{}
Iron and calcium contents, along with band centres and depths, are shown in Tab. \ref{vnirtab}.

For the majority of Vesta family members the calcium and iron content ($5<[Wo]<9$ and $31<[Fs]<40$) are compatible with an howarditic composition, in agreement with the results of De Sanctis et al. (2011b), although two objects (3155 and 3498) show a lower $[Wo]$ and $[Fs]$ content, compatible with a diogenite composition.
Non-vestoids show instead a  wider range of variation for calcium and iron molar contents, with one of the MOV (21238 Panarea) having the lowest molar content of the whole sample ($[Wo]  = 2.00$ and $[Fs] = 22.86$), while the IO 854 Frostia has the greatest iron/calcium content ($[Wo]  = 12.84$ and $[Fs] = 50.86$). The average values computed for  fugitives, low-i and NEAs (Tab. \ref{vnirtab}) are in agreement with an howarditic composition, although NEAs  show the greatest variation, having objects with both low and high iron content, which confirms the results we obtained on the basis of the NIR spectra.
No statistical conclusion was reached for IOs and MOVs, having only three and two objects in the VNIR subsample respectively, although  MOVs in particular show average values compatible with diogenites.

\section{Discussion and Conclusions}

We completed our statistical work over 117  V-type asteroids analyzed in the visible and/or near-infrared range in order to highlight similarities between the spectral characteristics of objects belonging to the Vesta dynamical family, V-type NEAs and several dynamical classes not connected to Vesta (fugitives, low-i, IOs, MOVs) called in literature ``non-vestoids''. Our statistical analysis has shown that:

\begin{itemize}
\item Two V-types (5379 and 6406), classified as basaltic objects in literature,  seem to belong to another taxonomic class.

\item V-type asteroids show greater visible slopes than HED meteorites. The  predominant lithology for basaltic objects is howardite, in agreement with the Dawn latest results, although some objects show an affinity with diogenites and eucrites. The analysis of band depth confirms that the majority of V-types are compatible with eucrites and howardites of a grain size $< 25 \, \mu $m.

 \item  Inner main belt V-type asteroids (fugitives, low-i, IOs)  have spectral parameters compatible with the Vesta family, pointing to Vesta as a plausible parent body, although  the  deconvolution of their spectra with the MGM show a great variation of their OPX/CPX content.

\item  NEAs and MOVs show spectral parameters quite different to the Vesta family. NEAs show also the greatest spread of iron content of the database, while the analysis on MOVs points to a different mineralogy. 

\end{itemize}

Laboratory experiments (Vernazza et al. 2006, Fulvio et al. 2012, 2015) have shown that space weathering affects the surfaces of basaltic material, reddening the spectral slope and lowering the band depths.
The higher spread of spectral parameters found on  NEAs could be
linked to a balance between space weathering processes and ``rejuvenation''  of surfaces caused by close encounters with planets (Binzel et al. 2010), which expose more fresh surface material (see Fulvio et al. 2015 for a detailed discussion). NEAs show also a higher spread of iron content  (Fig. \ref{nir}b) respect to the Vesta family.
Although we cannot exclude that in some cases the apparent diversity could be due to a different grain size (Fig. \ref{vnir}b),  our analysis  strongly support the idea that this diversity is due to a balance of weathering processes and a rejuvenation of surfaces triggered by close encounters with terrestrial planets.

The MOVs show higher band depths and in some cases  unusual band centres (Fig. \ref{vnir}). Their location in the main belt strongly points to a  different origin from Vesta since, as already pointed out, the probability that a 5 km V-type asteroid ejected from Vesta crosses the 3:1 resonance with Jupiter and reaches a stable orbit in the middle/outer belt, is almost 1\% (Roig et al. 2008). Hardersen et al. (2004) has already claimed for Magnya a different mineralogy.  In order to confirm or exclude a genetic link between Vesta and the MOV objects we looked for spots on the surface of Vesta having a composition and a mineralogy compatible with them. The comparison was performed using Dawn spectra collected by the VIR instrument on the south pole region, near the craters which likely produced the Vesta dynamical family.  
Comparing the maps of BI-BII centres and depths produced by the VIR Team (Ammannito et al. 2013) and the same parameters obtained using the same procedures for the two MOV objects in our sample, we found that Magnya and Panarea have spectral parameters not compatible with the south pole region of Vesta. Moreover, the excavation of a ($17 \pm 1$) km object like Magnya
 from the two deep craters around the south pole of Vesta (which have an estimated depth of 30-45 km) seems rather improbable. For these two basaltic objects, due to their peculiar spectral properties, sizes and location in the main belt we can argue that they  have an unrelated origin to Vesta. 

 These results were obtained for a small subsample of MOV objects. 
 In the near future  to enlarge the statistics for middle/outer V-type objects it will be necessary  to observe and spectrally confirm a large number of basaltic objects in the middle/outer main belt. The confirmation of a large clustered number of basaltic objects in a region dynamically inaccessible from Vesta would be the ironclad evidence of the existence of another basaltic parent body in the Solar System.

\section*{Acknowledgements}
In this article we make use of meteorite spectra taken with the NASA RELAB facility at Brown University.  This work was supported by INAF (PRIN-INAF 2011: Vesta as benchmark to understand Solar System history). We would like to thank Ren\'e Duffard and Maria Cristina De Sanctis for the useful discussions and the Dawn Team for the use of the VIR data.  We finally thank the referee Nicholas Moskovitz for the useful comments.






\appendix
\section{Tables}
%

{\scriptsize
       \begin{center}
     \begin{longtable} {|l|c|c|c|c|c|c|c|c|}  
\caption[]{Statistical sample: orbital elements (a, e, i), solar phase $\alpha$ and the author of visible and/or near-infrared observations and the dynamical group.} 
        \label{obs} \\
\hline \multicolumn{1}{|c|} {\textbf{Object}} &  \multicolumn{1}{c|} {\textbf{a}} & \multicolumn{1}{c|} {\textbf{e}} &    \multicolumn{1}{c|} {\textbf{sin i}}  &\multicolumn{1}{c|} {\textbf{$\alpha_v$}} & \multicolumn{1}{c|} {\textbf{author}} & \multicolumn{1}{c|} {\textbf{$\alpha_{NIR}$}} &    \multicolumn{1}{c|} {\textbf{author}}& \multicolumn{1}{c|} {\textbf{Group}}  \\  \hline 
\endfirsthead
\multicolumn{9}{c}%
{{\bfseries \tablename\ \thetable{} -- continued from previous page}} \\ \hline 
\endfoot
\hline \multicolumn{1}{|c|} {\textbf{Object}} &  \multicolumn{1}{c|} {\textbf{a}} & \multicolumn{1}{c|} {\textbf{e}} &    \multicolumn{1}{c|} {\textbf{sin i}}  &\multicolumn{1}{c|} {\textbf{$\alpha_v$}} & \multicolumn{1}{c|} {\textbf{author}} & \multicolumn{1}{c|} {\textbf{$\alpha_{NIR}$}} &    \multicolumn{1}{c|} {\textbf{author}}& \multicolumn{1}{c|} {\textbf{Group}}  \\  \hline 
\endhead
\hline \multicolumn{9}{r}{{Continue on next page}} \\ 
\endfoot
\hline \hline
\endlastfoot
809 Lundia	&	2.28	&	0.14	&	0.12	&	14.4	&	(1)	&	22.4	&	(4)	&	Fugitive	\\
	&		&		&		&		&		&	23.9	&	(10)	&		\\
854 Frostia	&	2.37	&	0.16	&	0.11	&	22.9	&	(2)	&	13.1	&	(3)	&	IO	\\
956 Elisa 	&	2.30	&	0.16	&	0.11	&	26.8	&	(1)	&	20.8	&	(4)	&	Fugitive	\\
 	&		&		&		&		&		&	16.6	&	(10)	&		\\
1459 Magnya	&	3.15	&	0.21	&	0.27	&	13.0	&	(1)	&	7.8	&	(11)	&	MOV	\\
1468 Zomba	&	2.18	&	0.21	&	0.18	&		&		&	13-38.3	&	(10)	&	IO	\\
1914 Hartbeespoortdam	&	2.41	&	0.14	&	0.08	&	12.9	&	(2)	&		&		&	Low-i	\\
$^{*}$1929 Kollaa&	2.36	&	0.11	&	0.12	&	3.8	&	(3)	&	14.4	&	(10)	&	Vestoid	\\
$^{*}$1933 Tinchen	&	2.35	&	0.09	&	0.12	&	11.9	&	(3)	&	7.9	&	(12)	&	Vestoid	\\
1946 Walraven	&	2.29	&	0.19	&	0.13	&	13.2	&	(2)	&		&		&	Fugitive	\\
1981 Midas	&	1.78	&	0.65	&	0.63	&	39.4	&	(3)	&	48	&	(3)	&	NEA	\\
$^{*}$2011 Veteraniya	&	2.39	&	0.11	&	0.11	&	27.9	&	(3)	&	16.8	&	(12)	&	Vestoid	\\
$^{*}$2045 Peking 	&	2.38	&	0.09	&	0.12	&	18.6	&	(3)	&	16.9-21.7	&	(10)	&	Vestoid	\\
 	&		&		&		&		&		&	13.9	&	(4)	&		\\
2371 Dimitrov	&	2.44	&	0.05	&	0.04	&	14.8	&	(3)	&	19.6	&	(10)	&	Low-i	\\
2442 Corbett	&	2.39	&	0.10	&	0.09	&	19.8	&	(3)	&	7.8	&	(10)	&	Low-i	\\
2468 Repin	&	2.33	&	0.12	&	0.11	&	12.3	&	(3)	&		&		&	Vestoid	\\
	&		&		&		&	21.4	&	 (4)	&	21.4	&	 (4)		&		\\
2486 Metsahovi	&	2.27	&	0.12	&	0.14	&	12.1	&	(2)	&		&		&	Fugitive	\\
2508 Alupka	&	2.37	&	0.09	&	0.11	&	11.5	&	(3)	&		&		&	Vestoid	\\
$^{*}$2511 Patterson	&	2.30	&	0.10	&	0.13	&	14.9	&	(3)	&	21.2	&	(10)	&	Vestoid	\\
2547 Hubei 	&	2.39	&	0.09	&	0.11	&	1.4	&	(3)	&		&		&	Vestoid	\\
2566 Kirghizia 	&	2.45	&	0.10	&	0.08	&	12.7	&	(3)	&	9.8	&	(10)	&	Low-i	\\
2579 Spartacus	&	2.21	&	0.08	&	0.11	&	10.8	&	(3)	&	18.9	&	(10)	&	Fugitive	\\
2640 Halllstrom	&	2.40	&	0.13	&	0.11	&	13.2	&	(3)	&		&		&	IO \\
2653 Principia	&	2.44	&	0.11	&	0.09	&	19.7	&	(3)	&	3.1-17.8	&	(10)	&	Low-i	\\
2704 Julian Loewe 	&	2.38	&	0.12	&	0.09	&	4.8	&	(3)	&		&		&	Low-i	\\
2763 Jeans	&	2.40	&	0.18	&	0.08	&	5.7	&	(3)	&	17.9	&	(4)	&	Low-i	\\
	&		&		&		&		&		&	12.2	&	(10)	&		\\
2795 Lepage	&	2.30	&	0.08	&	0.12	&	20.6	&	(3)	&	4.7	&	(10)	&	Fugitive	\\
2823 van der Laan	&	2.41	&	0.06	&	0.08	&		&		&	16.1	&	(10)	&	Low-i	\\
2851 Harbin	&	2.48	&	0.12	&	0.13	&	5.6	&	(3)	&	22.9	&	(4)	&	IO	\\
	&		&		&		&		&		&	8.7	&	 (10)	&		\\
2912 Lapalma	&	2.29	&	0.12	&	0.12	&	17.4	&	(3)	&	8.7	&	(10)	&	Fugitive	\\
$^{*}$3155 Lee	&	2.34	&	0.10	&	0.12	&	4.2-19.6	&	(3)	&	25.0	&	(4)	&	Vestoid	\\
 	&		&		&		&		&		&	7.9-8.4	&	(10)	&		\\
3265 Fletcher	&	2.41	&	0.11	&	0.11	&	3.4	&	(3)	&		&		&	Vestoid	\\
$^{*}$3268 De Sanctis	&	2.35	&	0.10	&	0.12	&	3.9	&	(3)	&	10	&	(4)	&	Vestoid	\\
3307 Athabasca	&	2.26	&	0.10	&	0.12	&	18.5	&	(3)	&		&		&	Vestoid	\\
$^{*}$3498 Belton	&	2.36	&	0.10	&	0.12	&	5.1	&	(3)	&	2.9	&	(4)	&	Vestoid	\\
3536 Schleicher	&	2.34	&	0.08	&	0.12	&	2.6	&	(3)	&		&		&	IO	\\
3613 Kunlun	&	2.37	&	0.12	&	0.12	&		&		&	9-18.5	&	(3)	&	Vestoid	\\
3657 Ermolova	&	2.31	&	0.09	&	0.12	&	11.7	&	(3)	&	25.4	&	(10)	&	Vestoid	\\
3703 Volkonskaya	&	2.33	&	0.09	&	0.12	&		&		&	9.8	&	(10)	&	Vestoid	\\
$^{*}$3782 Celle	&	2.42	&	0.11	&	0.11	&	10.6	&	(3)	&	23.2	&	(4)	&	Vestoid	\\
 	&		&		&		&		&		&	13.3-17.1	&	(10)	&		\\
3849 Incidentia	&	2.47	&	0.06	&	0.09	&	3.7	&	(3)	&		&		&	Low-i	\\
3850 Peltier	&	2.23	&	0.11	&	0.08	&	21.0	&	(3)	&		&		&	Fugitive	\\
3900 Knezevic	&	2.37	&	0.11	&	0.12	&	10.9	&	(3)	&		&		&	IO	\\
3908 Nyx	&	1.93	&	0.46	&	0.04	&	48.5	&	(3)	&	49.5	&	(13)	&	NEA	\\
$^{*}$3968 Koptelov	&	2.32	&	0.09	&	0.12	&	19.0	&	(3)	&	13.6	&	(12)	&	Vestoid	\\
$^{*}$4038 Kristina	&	2.37	&	0.10	&	0.11	&	4.8	&	(3)	&	10.3	&	(10)	&	Vestoid	\\
4055 Magellan	&	1.82	&	0.33	&	0.40	&	21.7	&	(1)	&	17.7	&	(13)	&	NEA	\\
4147 Lennon	&	2.36	&	0.10	&	0.11	&	15.2	&	(3)	&		&		&	Vestoid	\\
4188 Kitezh	&	2.34	&	0.11	&	0.10	&	0.4	&	(3)	&		&		&	Low-i	\\
$^{*}$4215 Kamo	&	2.42	&	0.10	&	0.12	&	16.7-25.3	&	(3)	&	3.2-17.3	&	(10)	&	Vestoid	\\
4278 Harvey	&	2.27	&	0.15	&	0.09	&	15.8	&	(1)	&		&		&	Fugitive	\\
4311 Zguridi	&	2.44	&	0.11	&	0.11	&	5.7	&	(3)	&		&		&	Vestoid	\\
4434 Nikulin	&	2.44	&	0.10	&	0.10	&	13.2	&	(3)	&		&		&	Low-i	\\
	&		&		&		&	11.5	&	(4)	&	11.5	&	(4)		&	\\
4796 Lewis	&	2.36	&	0.14	&	0.05	&	2.2	&	(3)	&	13.9	&	(10)	&	Low-i	\\
	&		&		&		&	7.8	&	(4)	&7.8	&	(4)		&		\\
4815 Anders 	&	2.36	&	0.10	&	0.13	&	3.4	&	(4)	&	3.4	&	(4)		&	Vestoid	\\
4900 Maymelou	&	2.38	&	0.10	&	0.11	&	21.8	&	(3)	&		&		&	Vestoid	\\
4977 Rauthgundis	&	2.29	&	0.09	&	0.10	&	17.7	&	(3)	&		&		&	Vestoid	\\
4993 Cossard	&	2.37	&	0.09	&	0.11	&	7.1	&	(3)	&		&		&	Vestoid	\\
$^{*}$5111 Jacliff	&	2.35	&	0.08	&	0.12	&	9.5	&	(3)	&	10.5	&	(10)	&	Vestoid	\\
5240 Kwasan	&	2.38	&	0.10	&	0.11	&	3.3	&	(3)	&		&		&	Vestoid	\\
5379 Abehiroshi	&	2.40	&	0.07	&	0.06	&	13.1	&	(3)	&	8.9	&	(3)	&	Low-i	\\
$^{*}$5481 Kiuchi	&	2.34	&	0.09	&	0.11	&	22.5	&	(1)	&	10.5	&	(10)	&	Vestoid	\\
5498 Gustafsson	&	2.25	&	0.10	&	0.05	&		&		&	7.8	&	(10)	&	Low-i	\\
5604 1992 FE	&	0.93	&	0.41	&	0.08	&	93.2	&	(3)	&	28.1	&	(13)	&	NEA	\\
6159 1991 YH	&	2.29	&	0.09	&	0.11	&	2.3	&	(4)	&	2.3	&	(4)		&	Vestoid	\\
6331 1992 FZ1	&	2.36	&	0.10	&	0.12	&		&		&	7.2	&	(4)	&	Vestoid	\\
6406 Vanavara	&	2.27	&	0.13	&	0.13	&	15.8	&	(2)	&	17.3	&	(12)	&	Fugitive	\\
6611 1993 VW	&	1.70	&	0.48	&	0.15	&	66.2	&	(3)	&	27.6	&	(13)	&	NEA	\\
7148 Reinholdbien	&	2.29	&	0.10	&	0.10	&	9.7	&	(2)	&	24.3	&	(12)	&	Fugitive	\\
7558 Yurlov	&	2.29	&	0.11	&	0.09	&	8.5	&	(5)	&		&		&	Fugitive	\\
7800 Zhongkeyuan	&	2.23	&	0.11	&	0.06	&		&		&	2.8	&	(10)	&	Low-i	\\
7889 1994 LX	&	1.26	&	0.35	&	0.60	&	43.7	&	(3)	&	21.9-25.1	&	(3)	&	NEA	\\
8693 Matsuki	&	2.41	&	0.12	&	0.11	&	7.7	&	(2)	&	13.8	&	(12)	&	Vestoid	\\
9481 Menchu	&	2.29	&	0.14	&	0.04	&		&		&	6.1	&	(10)	&	Low-i	\\
9553 Colas	&	2.20	&	0.11	&	0.04	&		&		&	20.3	&	(10)	&	Low-i	\\
10037 1984 BQ	&	2.39	&	0.10	&	0.12	&	4.2	&	(4)	& 4.2	&	(4)	&	Vestoid	\\
10285 Renemichelsen	&	2.35	&	0.10	&	0.12	&		&		& 5.8	&	(4)	&	Vestoid	\\
10320 Reiland	&	2.29	&	0.09	&	0.11	&		&		&6.5	&	(4)	&	Vestoid	\\
10349 1992 LN	&	2.38	&	0.10	&	0.12	&		&		&	12.7	&	(4)	&	Vestoid	\\
10537 1991 RY16	&	2.85	&	0.10	&	0.11	&	13.5-18.2	&	(5)	&		&		&	MOV	\\
	&		&		&		&	18.2	&	(6)	&		&		&		\\
$^{*}$11699 1998 FL105	&	2.40	&	0.11	&	0.10	&	1.1	&	(7)	&	2.4	&	(7)	&	Vestoid	\\
11876 Doncarpenter	&	2.43	&	0.12	&	0.11	&	5.2	&	(7)	&		&		&	Vestoid	\\
16416 1987 SM3	&	2.20	&	0.11	&	0.11	&		&		&	6	&	(10)	&	Fugitive	\\
16651 1993 TS11	&	2.29	&	0.10	&	0.10	&		&		&	5.5	&	(7)	&	Vestoid	\\
21238 Panarea	&	2.54	&	0.13	&	0.18	&	20.9	&	(8)	&	5.4	&	(3)	&	MOV	\\
22533 Krishnan	&	2.43	&	0.16	&	0.11	&	21.4	&	(7)	&		&		&	IO	\\
24941 1997 JM14	&	2.48	&	0.12	&	0.09	&	2.0	&	(5)	&		&		&	Low-i	\\
26886 1994 TJ2	&	2.34	&	0.12	&	0.08	&		&		&	10.9	&	(10)	&	Low-i	\\
27343 Deannashea	&	2.33	&	0.14	&	0.09	&		&		&	11.7	&	(10)	&	Low-i	\\
28517 2000 DD7	&	2.29	&	0.09	&	0.14	&	11.3	&	(5)	&		&		&	Fugitive	\\
33082 1997 WF43	&	2.30	&	0.08	&	0.11	&	13.3	&	(7)	&		&		&	Vestoid	\\
33881 2000 JK66	&	2.21	&	0.24	&	0.18	&		&		&	23.3	&	(10)	&	IO	\\
36412 2000 OP49	&	2.28	&	0.11	&	0.07	&		&		&	11.8	&	(10)	&	Low-i	\\
38070 Redwine	&	2.14	&	0.14	&	0.06	&	25.5	&	(5)	&	4.1	&	(10)	&	IO	\\
40521 1999 RL95	&	2.53	&	0.06	&	0.22	&	4.1	&	(8)	&		&		&	MOV	\\
42947 1999 TB98	&	2.41	&	0.09	&	0.11	&	17.8	&	(7)	&	12.1	&	(7)	&	Vestoid	\\
50098 2000 AG98	&	2.34	&	0.13	&	0.11	&		&		&	11.4	&	(10)	&	Vestoid	\\
52750 1998 KK17	&	1.47	&	0.53	&	0.19	&		&		&	21.0	&	(13)	&	NEA	\\
56570 2000 JA21	&	2.38	&	0.10	&	0.07	&	21.2	&	(5)	&		&		&	Low-i	\\
60669 2000 GE4	&	2.21	&	0.13	&	0.12	&	23.6	&	(5)	&		&		&	Fugitive	\\
66268 1999 JJ3	&	2.30	&	0.09	&	0.12	&	8.2	&	(7)	&	11.3	&	(7)	&	Vestoid	\\
88188 2000 XH44	&	2.01	&	0.39	&	0.20	&		&		&	9.6	&	(13)	&	NEA	\\
91290 1999 FR25	&	2.28	&	0.09	&	0.13	&	5.8	&	(7)	&		&		&	Vestoid	\\
97276 1999 XC143	&	2.49	&	0.15	&	0.06	&		&		&	2.4	&	(10)	&	Low-i	\\
137924 2000BD19	&	0.88	&	0.90	&	0.43	&		&		&	19.9	&	(3)	&	NEA	\\
192563 1998WZ6	&	1.45	&	0.41	&	0.42	&		&		&	27.4	&	(3)	&	NEA	\\
238063 2003 EG	&	1.74	&	0.71	&	0.53	&	45.6	&	(9)	&		&		&	NEA	\\
253841 2003YG118	&	2.29	&	0.64	&	0.14	&		&		&	48.2	&	(3)	&	NEA	\\
297418 2000SP43	&	0.81	&	0.47	&	0.18	&		&		&	35.7-51.5	&	(3)	&	NEA	\\
326290 Akhenaten	&	0.88	&	0.44	&	0.06	&		&		&	14.5	&	(3)	&	NEA	\\
2001YE4	&	0.68	&	0.54	&	0.08	&		&		&	57.2	&	(3)	&	NEA	\\
2003 FU3	&	0.86	&	0.39	&	0.23	&	67.6	&	(9)	&		&		&	NEA	\\
2003 FT3	&	2.68	&	0.57	&	0.07	&	21.8	&	(9)	&	47.4	&	(3)	&	NEA	\\
2003 GJ21	&	1.81	&	0.40	&	0.12	&	16.9	&	(9)	&		&		&	NEA	\\
2005 WX	&	1.60	&	0.38	&	0.08	&		&		&	29.9	&	(13)	&	NEA	\\
2008 BT18	&	2.22	&	0.59	&	0.14	&		&		&	34.4	&		(14)&	NEA	\\
2011YA	&	2.12	&	0.76	&	0.09	&		&		&	22.7	&	(3)	&	NEA	\\
2013KL6	&	2.10	&	0.54	&	0.13	&		&		&	0.4	&	(3)	&	NEA	\\
\hline
\hline
\end{longtable}

\end{center}
Spectra were taken from: (1) the S3OS2 survey (Lazzaro et al. 2004); (2) Alvarez-Candal et al. (2006); (3) the SMASS survey\footnote{http://smass.mit.edu/smass.html}; (4) Duffard et al. (2004); (5) Moskovitz et al.  (2008a, 2008b); (6) Duffard \& Roig (2009); (7) Jasmim et al. (2013); (8) Roig et al. (2008); (9) Marchi et al. (2005); (10) Moskovitz et al. (2010); (11) Hardersen et al. (2004); (12) De Sanctis et al. (2011b); (13)  Burbine et al. (2009); (14) Reddy, Emery \& Gaffey (2008). Objects belonging to the control sample are marked with an asterisk.}

{\scriptsize
       \begin{center}
     \begin{longtable} {|l|c|c|c|c|}  
\caption[]{Visible spectral analysis: SlopeA, Apparent Depth, Slope B and SlopeA after the empirical correction by Reddy et al. (2012).}. 
        \label{vistab} \\
\hline \multicolumn{1}{|c|} {\textbf{Object}} & \multicolumn{1}{c|}
{\textbf{SlopeA ($\%/10^3 \textup{\AA}$)}}  &
\multicolumn{1}{c|} {\textbf{Apparent Depth}} &     \multicolumn{1}{c|}
{\textbf{SlopeB ($\%/10^3 \textup{\AA}$)}} & \multicolumn{1}{c|}
{\textbf{SlopeAcorr ($\%/10^3 \textup{\AA}$)}}  \\  \hline
\endfirsthead
\multicolumn{4}{c}%
{{\bfseries \tablename\ \thetable{} -- continued from previous page}} \\ \hline 
\endfoot
\hline \multicolumn{1}{|c|} {\textbf{Object}} & \multicolumn{1}{c|}
{\textbf{SlopeA ($\%/10^3 \textup{\AA}$)}} & 
\multicolumn{1}{c|} {\textbf{Apparent Depth}} &     \multicolumn{1}{c|}
{\textbf{SlopeB ($\%/10^3 \textup{\AA}$)}}  & \multicolumn{1}{c|}
{\textbf{SlopeAcorr ($\%/10^3 \textup{\AA}$)}} \\  \hline 
\endhead
\hline \multicolumn{4}{r}{{Continue on next page}} \\ 
\endfoot
\hline \hline
\endlastfoot 
\hline
809	Lundia	&	15.24	$\pm$	0.09	&	1.68	$\pm$	0.10	&	-32.07	$\pm$	0.46	&		10.39 	\\
854	Frostia	&			-------------	&	1.41	$\pm$	0.20	&	-31.26	$\pm$	1.24	&		12.07	\\
956	Elisa	&	16.34	$\pm$	0.11	&	1.67	$\pm$	0.03	&	-30.77	$\pm$	0.40	&		12.84	\\
1459	Magnya	&	15.55	$\pm$	0.16	&	1.81	$\pm$	0.06	&	-32.48	$\pm$	0.60	&		10.11	\\
1914	Hartbeespoortdam	&			-------------	&	1.41	$\pm$	0.04	&	-20.83	$\pm$	0.27	&		10.09	\\
1929	Kollaa	&	9.83	$\pm$	0.11	&	1.54	$\pm$	0.04	&	-25.02	$\pm$	0.52	&		8.29	\\
1933	Tinchen	&	12.83	$\pm$	0.16	&	1.35	$\pm$	0.01	&	-22.44	$\pm$	0.37	&		9.89	\\
1946	Walraven	&			-------------	&	1.50	$\pm$	0.03	&	-27.63	$\pm$	0.30	&		10.15	\\
1981	Midas	&	9.03	$\pm$	0.35	&	1.62	$\pm$	0.07	&	-22.83	$\pm$	0.96	&			\\
2011	Veteraniya	&	14.82	$\pm$	0.19	&	1.48	$\pm$	0.12	&	-24.20	$\pm$	1.02	&		13.06	\\
2045	Peking	&	12.88	$\pm$	0.24	&	1.48	$\pm$	0.08	&	-24.42	$\pm$	0.97	&		11.22	\\
2371	Dimitrov	&	14.10	$\pm$	0.18	&	1.41	$\pm$	0.08	&	-21.81	$\pm$	1.14	&		10.47	\\
2442	Corbett	&	14.56	$\pm$	0.31	&	1.62	$\pm$	0.06	&	-28.92	$\pm$	1.52	&		11.46	\\
2468	Repin	&	11.65	$\pm$	0.14	&	1.61	$\pm$	0.17	&	-29.28	$\pm$	0.55	&		9.97	\\
2486	Metsahovi	&			-------------	&	1.59	$\pm$	0.07	&	-19.17	$\pm$	0.30	&		9.93	\\
2508	Alupka	&	10.58	$\pm$	0.21	&	1.45	$\pm$	0.09	&	-25.31	$\pm$	1.37	&		9.82	\\
2511	Patterson	&	12.90	$\pm$	0.29	&	1.55	$\pm$	0.07	&	-25.89	$\pm$	0.97	&		10.49	\\
2547	Hubei	&	9.99	$\pm$	0.22	&	1.42	$\pm$	0.11	&	-20.46	$\pm$	0.97	&		7.82	\\
2566	Kirghizia	&	11.95	$\pm$	0.20	&	1.57	$\pm$	0.08	&	-25.24	$\pm$	1.06	&		10.05	\\
2579	Spartacus	&	15.76	$\pm$	0.43	&	1.90	$\pm$	0.07	&	-38.64	$\pm$	1.31	&		9.68	\\
2640	Halllstrom	&	10.89	$\pm$	0.17	&	1.45	$\pm$	0.08	&	-22.21	$\pm$	1.07	&		10.15	\\
2653	Principia	&	14.34	$\pm$	0.25	&	1.72	$\pm$	0.08	&	-31.48	$\pm$	1.02	&		11.44	\\
2704	Julian Loewe	&	10.51	$\pm$	0.30	&	1.54	$\pm$	0.03	&	-22.60	$\pm$	1.00	&		8.49	\\
2763	Jeans	&	12.63	$\pm$	0.22	&	1.50	$\pm$	0.10	&	-26.45	$\pm$	1.31	&		8.67	\\
2795	Lepage	&	15.01	$\pm$	0.33	&	1.63	$\pm$	0.07	&	-27.87	$\pm$	0.90	&		11.62	\\
2851	Harbin	&	19.53	$\pm$	0.43	&	1.74	$\pm$	0.07	&	-30.18	$\pm$	1.66	&		8.65	\\
2912	Lapalma	&	14.73	$\pm$	0.31	&	1.72	$\pm$	0.06	&	-25.07	$\pm$	1.21	&		10.98	\\
3155	Lee	&	15.65	$\pm$	0.28	&	1.60	$\pm$	0.06	&	-26.83	$\pm$	0.86	&		8.37	\\
3265	Fletcher	&	7.93	$\pm$	0.23	&	1.33	$\pm$	0.03	&	-21.80	$\pm$	0.77	&		8.21	\\
3268	De Sanctis	&	9.95	$\pm$	0.19	&	1.31	$\pm$	0.05	&	-19.89	$\pm$	0.88	&		8.31	\\
3307	Athabasca	&	15.44	$\pm$	0.35	&	1.85	$\pm$	0.09	&	-27.44	$\pm$	1.28	&		11.20	\\
3498	Belton	&	7.51	$\pm$	0.25	&	1.35	$\pm$	0.16	&	-22.61	$\pm$	1.36	&		8.55	\\
3536	Schleicher	&	9.60	$\pm$	0.20	&	1.38	$\pm$	0.05	&	-22.89	$\pm$	1.15	&		8.05	\\
3657	Ermolova	&	12.79	$\pm$	0.18	&	1.56	$\pm$	0.03	&	-23.80	$\pm$	0.46	&		9.85	\\
3782	Celle	&	11.37	$\pm$	0.31	&	1.51	$\pm$	0.13	&	-25.51	$\pm$	0.93	&		9.64	\\
3849	Incidentia	&	10.30	$\pm$	0.27	&	1.54	$\pm$	0.07	&	-27.03	$\pm$	1.14	&		8.27	\\
3850	Peltier	&	12.88	$\pm$	0.31	&	1.65	$\pm$	0.10	&	-26.54	$\pm$	0.80	&		11.70	\\
3900	Knezevic	&	12.78	$\pm$	0.26	&	1.62	$\pm$	0.03	&	-24.76	$\pm$	0.92	&		9.70	\\
3908	Nyx	&	9.87	$\pm$	0.32	&	1.87	$\pm$	0.01	&	-23.40	$\pm$	1.03	&		\\
3968	Koptelov	&	9.98	$\pm$	0.14	&	1.32	$\pm$	0.07	&	-21.47	$\pm$	0.70	&		11.30	\\
4038	Kristina	&	13.04	$\pm$	0.14	&	1.21	$\pm$	0.06	&	-19.00	$\pm$	0.59	&		8.49	\\
4055	Magellan	&	12.48	$\pm$	0.28	&	1.52	$\pm$	0.14	&	-18.79	$\pm$	1.54	&	\\
4147	Lennon	&	8.54	$\pm$	0.18	&	1.31	$\pm$	0.02	&	-18.70	$\pm$	0.72	&		10.55	\\
4188	Kitezh	&	9.97	$\pm$	0.25	&	1.46	$\pm$	0.01	&	-20.15	$\pm$	1.11	&		7.62	\\
4215	Kamo	&	11.84	$\pm$	0.21	&	1.53	$\pm$	0.02	&	-27.61	$\pm$	0.70	&		10.84	\\
4278	Harvey	&	11.78	$\pm$	0.10	&	1.46	$\pm$	0.08	&	-25.62	$\pm$	0.67	&		10.67	\\
4311	ZguridI	&	9.12	$\pm$	0.50	&	1.35	$\pm$	0.15	&	-16.96	$\pm$	1.95	&		8.67	\\
4434	Nikulin	&	9.91	$\pm$	0.19	&	1.49	$\pm$	0.06	&	-26.89	$\pm$	0.65	&		10.15	\\
4796	Lewis	&	10.49	$\pm$	0.30	&	1.48	$\pm$	0.11	&	-22.02	$\pm$	1.27	&		7.97	\\
4815	Anders	&	14.51	$\pm$	0.16	&	1.26	$\pm$	0.09	&	-21.32	$\pm$	0.72	&		8.31	\\
4900	Maymelou	&	12.33	$\pm$	0.19	&	1.51	$\pm$	0.10	&	-25.75	$\pm$	0.94	&		11.85	\\
4977	Rauthgundis	&	11.88	$\pm$	0.32	&	1.67	$\pm$	0.09	&	-19.97	$\pm$	1.71	&		11.04	\\
4993	Cossard	&	10.53	$\pm$	0.33	&	1.47	$\pm$	0.11	&	-22.88	$\pm$	1.00	&		8.94	\\
5111	Jacliff	&	13.35	$\pm$	0.33	&	1.42	$\pm$	0.01	&	-22.12	$\pm$	1.66	&		9.42	\\
5240	Kwasan	&	10.00	$\pm$	0.19	&	1.49	$\pm$	0.04	&	-24.54	$\pm$	0.99	&		8.19	\\
5379	Abehiroshi	&	10.04	$\pm$	0.64	&	1.74	$\pm$	0.10	&	-39.51	$\pm$	2.50	&		10.13	\\
5481	Kiuchi	&	18.32	$\pm$	0.19	&	1.69	$\pm$	0.22	&	-35.40	$\pm$	0.58	&		11.99	\\
5604	1992 FE	&	14.71	$\pm$	0.32	&	2.06	$\pm$	1.14	&	-40.57	$\pm$	3.75	&	\\
6159	1991 YH	&	14.38	$\pm$	0.15	&	1.55	$\pm$	0.27	&	-23.31	$\pm$	0.70	&		7.89	\\
6406	Vanavara	&	11.82	$\pm$	0.07	&	1.25	$\pm$	0.08	&	-12.27	$\pm$	0.63	&		10.67	\\
6611	1993 VW	&	11.81	$\pm$	0.29	&	1.63	$\pm$	0.20	&	-29.95	$\pm$	1.72	&	\\
7148	Reinholdbien	&	12.56	$\pm$	0.06	&	1.51	$\pm$	0.10	&	-23.80	$\pm$	0.57	&		9.46	\\
7558	Yurlov	&	13.05	$\pm$	0.24	&	1.48	$\pm$	0.06	&	-28.11	$\pm$	0.73	&		9.22	\\
7889	1994 LX	&	8.96	$\pm$	0.71	&	2.08	$\pm$	0.47	&	-29.85	$\pm$	2.80	&	\\
8693	Matsuki&				-------------	&	1.24	$\pm$	0.03	&	-15.19	$\pm$	0.26	&		9.06	\\
10037	1984 BQ	&	13.83	$\pm$	0.08	&	1.42	$\pm$	0.06	&	-26.91	$\pm$	0.35	&		8.25	\\
10537	1991 RY16	&	8.79	$\pm$	0.39	&	1.50	$\pm$	0.01	&	-28.23	$\pm$	0.58	&		10.29	\\
11699	1998 FL105	&	7.59	$\pm$	0.30	&	1.63	$\pm$	0.07	&	-26.59 $\pm$	2.72		&		7.76	\\
11876	Doncarpenter	&	8.24	$\pm$	0.17	&	-------------	&	-------------						&	8.57	\\
21238	Panarea	&	10.09	$\pm$	0.15	&	1.80	$\pm$	0.55	&	-23.83	$\pm$	1.30	&		11.68	\\
22533	Krishnan	&	6.49	$\pm$	0.80	&	-------------	&	-------------						&	11.78	\\
24941	1997 JM14	&	10.86	$\pm$	0.15	&	1.34	$\pm$	0.01	&	-25.72	$\pm$	0.67	&		7.93	\\
28517	2000DD7	&	11.47	$\pm$	0.16	&	1.50	$\pm$	0.03	&	-28.58	$\pm$	0.79	&		9.78	\\
33082	1997 WF43	&	8.71	$\pm$	0.23	&	-------------	&	-------------						&	10.17	\\
38070	Redwine	&	14.79	$\pm$	0.25	&	1.55	$\pm$	0.07	&	-26.22	$\pm$	1.36	&		12.59	\\
40521	1999 RL95	&	9.99	$\pm$	0.19	&	1.78	$\pm$	0.59	&	-26.55	$\pm$	1.66	&		8.35	\\
42947	1999 TB98	&	7.87	$\pm$	0.16	&	1.66	$\pm$	0.06	&	-23.02 $\pm$	2.59		&		11.06	\\
56570	2000 JA21	&	14.68	$\pm$	0.24	&	1.62	$\pm$	0.17	&	-27.31	 $\pm$	1.86	&		11.74	\\
60669	2000 GE4	&	13.30	$\pm$	0.29	&	1.73	$\pm$	0.16	&	-28.45	$\pm$	1.43	&		12.21	\\
66268	1999 JJ3	&	12.25	$\pm$	0.19	&	1.44	$\pm$	0.03	&	 -21.89	$\pm$ 3.79		&		9.16	\\
91290	1999 FR25	&	8.67	$\pm$	0.37	&	-------------	&	-------------						&	8.69	\\
238063	2003EG	&	5.87	$\pm$	0.26	&	1.18	$\pm$	0.18	&	-12.73	$\pm$	1.38	&		\\
2003	FT3	&	5.62	$\pm$	0.23	&	1.18	$\pm$	0.10	&	-12.96	$\pm$	1.48	&		\\
2003	FU3	&	3.14	$\pm$	0.24	&	1.10	$\pm$	0.07	&	-8.02	$\pm$	0.88	&		\\
2003	GJ21	&	3.15	$\pm$	0.1	&	1.10	$\pm$	0.04	&	-6.11	$\pm$	0.45	&		\\
																	
\hline																	
\hline																	
Vestoids		&	11.51	$\pm$	2.73	&	1.48	$\pm$	0.13	&	-23.82	$\pm$	3.51	&			\\
Fugitives		&	13.81	$\pm$	1.66	&	1.62	$\pm$	0.13	&	-27.87	$\pm$	4.56	&			\\
IOs		&	12.35	$\pm$	4.51	&	1.53	$\pm$	0.14	&	-26.25	$\pm$	3.75	&			\\
Low-i		&	12.06	$\pm$	1.90	&	1.52	$\pm$	0.10	&	-25.11	$\pm$	3.40	&			\\
NEAs		&	8.46	$\pm$	3.95	&	1.53	$\pm$	0.38	&	-20.52	$\pm$	10.94	&			\\
MOVs		&	11.07	$\pm$	3.06	&	1.72	$\pm$	0.15	&	-27.77	$\pm$	3.62	&			\\
Control	sample	&	12.12	$\pm$	2.92	&	1.46	$\pm$	0.13	&	-24.60	$\pm$	3.92	&			\\							
\end{longtable}
\end{center}
}

{\scriptsize
       \begin{center}
     \begin{longtable} {|l|c|c|c|c|}  
\caption[]{Visible spectral analysis: SlopeA, Apparent Depth and Slope B for a HED meteorite sample}. 
        \label{vished} \\
        \hline
\textbf{Sample name} 	&	\textbf{SubType} &	\textbf{SlopeA ($\%/10^3 \textup{\AA}$)}		&	\textbf{Apparent Depth} 		&	\textbf{SlopeB ($\%/10^3 \textup{\AA}$)}		\\ \hline \hline
A-881526	&	Diogenite	&	9.70	$\pm$	0.86	&	2.58	$\pm$	0.10	&	-36.46	$\pm$	1.95	\\
Aioun el Atrouss	&	Diogenite	&	4.65	$\pm$	0.35	&	2.45	$\pm$	0.10	&	-35.34	$\pm$	1.77	\\
EETA79002	&	Diogenite	&	3.08	$\pm$	0.23	&	1.89	$\pm$	0.05	&	-33.66	$\pm$	1.43	\\
GRO95555	&	Diogenite	&	6.90	$\pm$	0.77	&	2.37	$\pm$	0.09	&	-36.30	$\pm$	1.79	\\
Johnstown	&	Diogenite	&	4.65	$\pm$	0.35	&	1.74	$\pm$	0.04	&	-30.55	$\pm$	1.31	\\
LAP91900	&	Diogenite	&	5.67	$\pm$	0.41	&	2.44	$\pm$	0.10	&	-36.85	$\pm$	1.84	\\
Roda	&	Diogenite	&	4.98	$\pm$	0.39	&	2.30	$\pm$	0.07	&	-33.99	$\pm$	1.88	\\
Shalka	&	Diogenite	&	3.88	$\pm$	0.35	&	2.30	$\pm$	0.05	&	-33.27	$\pm$	2.02	\\
Tatahouine	&	Diogenite	&	1.39	$\pm$	0.52	&	2.39	$\pm$	0.09	&	-33.24	$\pm$	1.71	\\
Y-74013	&	Diogenite	&	5.98	$\pm$	0.53	&	1.63	$\pm$	0.04	&	-31.32	$\pm$	1.04	\\
Y-75032	&	Diogenite	&	5.59	$\pm$	0.29	&	1.89	$\pm$	0.05	&	-36.45	$\pm$	1.30	\\
\hline
Binda	&	Howardite	&	7.10	$\pm$	0.33	&	2.57	$\pm$	0.10	&	-38.90	$\pm$	2.02	\\
Bununu	&	Howardite	&	5.16	$\pm$	0.21	&	1.68	$\pm$	0.04	&	-31.96	$\pm$	0.97	\\
EET83376	&	Howardite	&	10.54	$\pm$	0.18	&	1.66	$\pm$	0.04	&	-37.89	$\pm$	1.07	\\
EET87503	&	Howardite	&	7.39	$\pm$	0.11	&	1.40	$\pm$	0.03	&	-28.04	$\pm$	0.60	\\
EET87513	&	Howardite	&	6.93	$\pm$	0.25	&	1.46	$\pm$	0.03	&	-29.35	$\pm$	0.70	\\
Frankfort	&	Howardite	&	4.13	$\pm$	0.13	&	1.90	$\pm$	0.05	&	-37.29	$\pm$	1.36	\\
GRO95535	&	Howardite	&	6.91	$\pm$	0.21	&	1.62	$\pm$	0.03	&	-32.75	$\pm$	0.97	\\
GRO95574	&	Howardite	&	5.01	$\pm$	0.10	&	1.53	$\pm$	0.03	&	-29.77	$\pm$	0.82	\\
Kapoeta	&	Howardite	&	4.05	$\pm$	0.23	&	1.55	$\pm$	0.03	&	-29.10	$\pm$	0.80	\\
Le Teilleul	&	Howardite	&	7.44	$\pm$	0.16	&	2.04	$\pm$	0.06	&	-40.42	$\pm$	1.64	\\
Pavlovka	&	Howardite	&	4.78	$\pm$	0.27	&	1.97	$\pm$	0.09	&	-30.86	$\pm$	1.80	\\
Petersburg	&	Howardite	&	8.81	$\pm$	0.23	&	2.14	$\pm$	0.06	&	-37.67	$\pm$	1.86	\\
QUE94200	&	Howardite	&	6.38	$\pm$	0.25	&	1.80	$\pm$	0.04	&	-35.15	$\pm$	1.34	\\
Y-7308	&	Howardite	&	5.20	$\pm$	0.31	&	2.16	$\pm$	0.07	&	-37.51	$\pm$	1.63	\\
Y-790727	&	Howardite	&	14.13	$\pm$	0.51	&	1.76	$\pm$	0.04	&	-40.26	$\pm$	1.37	\\
Y-791573	&	Howardite	&	15.44	$\pm$	0.62	&	1.75	$\pm$	0.03	&	-39.33	$\pm$	1.39	\\
\hline
A-881819	&	Eucrite	&	4.93	$\pm$	0.28	&	1.77	$\pm$	0.10	&	-33.37	$\pm$	1.06	\\
ALH-78132	&	Eucrite	&	10.23	$\pm$	0.09	&	1.59	$\pm$	0.04	&	-35.83	$\pm$	1.06	\\
ALHA76005	&	Eucrite	&	10.25	$\pm$	0.28	&	1.62	$\pm$	0.06	&	-35.59	$\pm$	1.92	\\
ALHA81001	&	Eucrite	&	7.15	$\pm$	0.17	&	1.31	$\pm$	0.06	&	-27.91	$\pm$	0.49	\\
Bereba	&	Eucrite	&	4.32	$\pm$	0.25	&	1.59	$\pm$	0.05	&	-23.66	$\pm$	1.27	\\
Bouvante	&	Eucrite	&	5.96	$\pm$	0.61	&	1.55	$\pm$	0.10	&	-36.05	$\pm$	1.36	\\
EETA79005	&	Eucrite	&	12.04	$\pm$	0.20	&	1.68	$\pm$	0.04	&	-38.98	$\pm$	1.13	\\
Ibitira	&	Eucrite	&	3.67	$\pm$	0.27	&	1.48	$\pm$	0.07	&	-30.52	$\pm$	1.06	\\
Juvinas	&	Eucrite	&	6.50	$\pm$	0.14	&	1.71	$\pm$	0.04	&	-35.69	$\pm$	1.15	\\
LEW85303	&	Eucrite	&	9.47	$\pm$	0.19	&	1.51	$\pm$	0.07	&	-34.06	$\pm$	0.68	\\
LEW87004	&	Eucrite	&	9.43	$\pm$	0.27	&	1.66	$\pm$	0.07	&	-33.08	$\pm$	1.66	\\
Millbillillie	&	Eucrite	&	4.11	$\pm$	0.10	&	1.40	$\pm$	0.02	&	-24.95	$\pm$	0.56	\\
Nobleborough	&	Eucrite	&	10.58	$\pm$	0.17	&	1.68	$\pm$	0.07	&	-32.10	$\pm$	1.46	\\
Pasamonte	&	Eucrite	&	11.18	$\pm$	0.15	&	1.56	$\pm$	0.07	&	-30.57	$\pm$	1.07	\\
PCA82501	&	Eucrite	&	7.72	$\pm$	0.16	&	1.38	$\pm$	0.03	&	-28.41	$\pm$	0.59	\\
PCA82502	&	Eucrite	&	10.56	$\pm$	0.25	&	1.60	$\pm$	0.10	&	-40.60	$\pm$	0.92	\\
PCA91007	&	Eucrite	&	7.31	$\pm$	0.44	&	1.65	$\pm$	0.09	&	-36.84	$\pm$	1.28	\\
Serra de Mage	&	Eucrite	&	3.09	$\pm$	0.22	&	1.67	$\pm$	0.04	&	-26.95	$\pm$	1.08	\\
Sioux County	&	Eucrite	&	4.27	$\pm$	0.10	&	1.45	$\pm$	0.04	&	-21.80	$\pm$	0.99	\\
Stannern	&	Eucrite	&	7.42	$\pm$	0.17	&	1.49	$\pm$	0.04	&	-27.24	$\pm$	1.20	\\
Y-74450	&	Eucrite	&	9.06	$\pm$	0.12	&	1.50	$\pm$	0.06	&	-31.28	$\pm$	0.79	\\
Y-792769	&	Eucrite	&	9.00	$\pm$	0.12	&	1.49	$\pm$	0.08	&	-34.31	$\pm$	0.60	\\
Y-793591	&	Eucrite	&	9.77	$\pm$	0.10	&	1.51	$\pm$	0.04	&	-35.86	$\pm$	0.75	\\
Y-82082	&	Eucrite	&	12.31	$\pm$	0.46	&	1.47	$\pm$	0.07	&	-33.07	$\pm$	0.64	\\
\hline \hline
\end{longtable}
\end{center}
}

{\scriptsize
       \begin{center}
     \begin{longtable} {|l|c|c|c|}  
\caption[]{Infrared spectral analysis: BI minimum, BII minimum and Band Separation}. 
        \label{nirtab} \\
\hline \multicolumn{1}{|c|} {\textbf{Object}} & \multicolumn{1}{c|}
{\textbf{BI minimum ($\mu m$)}} & \multicolumn{1}{c|} {\textbf{BII minimum ($\mu m$)}} &
\multicolumn{1}{c|} {\textbf{Band separation ($\mu m$)}} \\  \hline 
\endfirsthead
\multicolumn{4}{c}%
{{\bfseries \tablename\ \thetable{} -- continue on next page}} \\ \hline 
\endfoot
\hline \multicolumn{1}{|c|} {\textbf{Object}} & \multicolumn{1}{c|}
{\textbf{BI minimum ($\mu m$)}} & \multicolumn{1}{c|} {\textbf{BII minimum ($\mu m$)}} &
\multicolumn{1}{c|} {\textbf{Band separation ($\mu m$)}} \\   \hline   
\endhead
\hline \multicolumn{4}{r}{{Continue on next page}} \\ 
\endfoot
\hline \hline
\endlastfoot
809 Lundia	&	0.921	$\pm$	0.002	&	1.939	$\pm$	0.001	&	1.02	$\pm$	0.01	\\
854 Frostia	&	0.950	$\pm$	0.001	&	2.000	$\pm$	0.002	&	1.05	$\pm$	0.01	\\
956 Elisa 	&	0.922	$\pm$	0.002	&	1.929	$\pm$	0.002	&	1.01	$\pm$	0.01	\\
1459 Magnya	&	0.928	$\pm$	0.001	&	1.923	$\pm$	0.001	&	1.00	$\pm$	0.01	\\
1468 Zomba	&	0.923	$\pm$	0.002	&	1.973	$\pm$	0.003	&	1.05	$\pm$	0.01	\\
1929 Kollaa	&	0.927	$\pm$	0.003	&	1.957	$\pm$	0.003	&	1.03	$\pm$	0.01	\\
1933 Tinchen	&	0.916	$\pm$	0.008	&	1.939	$\pm$	0.002	&	1.02	$\pm$	0.01	\\
1981 Midas	&	0.930	$\pm$	0.002	&	2.020	$\pm$	0.011	&	1.09	$\pm$	0.01	\\
2011 Veteraniya	&	0.916	$\pm$	0.002	&	1.952	$\pm$	0.006	&	1.04	$\pm$	0.01	\\
2045 Peking 	&	0.926	$\pm$	0.002	&	1.939	$\pm$	0.002	&	1.01	$\pm$	0.01	\\
2371 Dimitrov	&	0.920	$\pm$	0.002	&	1.972	$\pm$	0.006	&	1.05	$\pm$	0.01	\\
2442 Corbett	&	0.922	$\pm$	0.001	&	1.935	$\pm$	0.001	&	1.01	$\pm$	0.01	\\
2511 Patterson	&	0.925	$\pm$	0.002	&	1.955	$\pm$	0.004	&	1.03	$\pm$	0.01	\\
2566 Kirghizia 	&	0.925	$\pm$	0.002	&	1.958	$\pm$	0.006	&	1.03	$\pm$	0.01	\\
2579 Spartacus  	&	0.935	$\pm$	0.004	&	1.990	$\pm$	0.012	&	1.06	$\pm$	0.02	\\
2653 Principia 	&	0.925	$\pm$	0.001	&	1.964	$\pm$	0.007	&	1.04	$\pm$	0.01	\\
2763 Jeans	&	0.930	$\pm$	0.002	&	1.995	$\pm$	0.003	&	1.07	$\pm$	0.01	\\
2795 Lepage	&	0.928	$\pm$	0.002	&	1.950	$\pm$	0.003	&	1.02	$\pm$	0.01	\\
2823 van der Laan	&	0.925	$\pm$	0.001	&	1.953	$\pm$	0.005	&	1.03	$\pm$	0.01	\\
2851 Harbin	&	0.920	$\pm$	0.001	&	1.907	$\pm$	0.004	&	0.99	$\pm$	0.01	\\
2912 Lapalma	&	0.920	$\pm$	0.001	&	1.927	$\pm$	0.003	&	1.01	$\pm$	0.01	\\
3155 Lee	&	0.910	$\pm$	0.002	&	1.913	$\pm$	0.007	&	1.00	$\pm$	0.01	\\
3268 De Sanctis &	0.907	 $\pm$	0.005&	1.988 $\pm$		0.016	&1.08 $\pm$	0.02\\
3498 Belton	&	0.917	$\pm$	0.006	&	1.889	$\pm$	0.021	&	0.97	$\pm$	0.03	\\
3613 Kunlun	&	0.960	$\pm$	0.004	&	1.943	$\pm$	0.006	&	0.98	$\pm$	0.01	\\
3657 Ermolova	&	0.921	$\pm$	0.001	&	1.919	$\pm$	0.003	&	1.00	$\pm$	0.01	\\
3703 Volkonskaya	&	0.915	$\pm$	0.004	&	1.932	$\pm$	0.007	&	1.02	$\pm$	0.01	\\
3782 Celle 	&	0.920	$\pm$	0.002	&	1.942	$\pm$	0.004	&	1.02	$\pm$	0.01	\\
3908 Nyx	&	0.923	$\pm$	0.003	&	1.943	$\pm$	0.004	&	1.02	$\pm$	0.01	\\
3968 Koptelov	&	0.922	$\pm$	0.009	&	1.905	$\pm$	0.001	&	0.98	$\pm$	0.01	\\
4038 Kristina	& 0.910$\pm$		0.001&	1.955	$\pm$	0.002&	1.05 $\pm$	0.01\\
4055 Magellan	&	0.920	$\pm$	0.001	&	1.925	$\pm$	0.003	&	1.01	$\pm$	0.01	\\
4215 Kamo	&	0.920	$\pm$	0.002	&	1.945	$\pm$	0.014	&	1.03	$\pm$	0.02	\\
4796 Lewis	&	0.925	$\pm$	0.002	&	1.951	$\pm$	0.008	&	1.03	$\pm$	0.01	\\
5111 Jacliff	&	0.920	$\pm$	0.002	&	1.940	$\pm$	0.001	&	1.02	$\pm$	0.01	\\
5481 Kiuchi	&	0.919	$\pm$	0.002	&	1.939	$\pm$	0.008	&	1.02	$\pm$	0.01	\\
5498 Gustafsson	&	0.928	$\pm$	0.001	&	1.967	$\pm$	0.002	&	1.04	$\pm$	0.01	\\
5604 1992 FE	&	0.918	$\pm$	0.002	&	1.953	$\pm$	0.009	&	1.04	$\pm$	0.01	\\
6331 1992 FZ1	&	0.901	$\pm$	0.002	&	2.054	$\pm$	0.002	&	1.15	$\pm$	0.01	\\
6611 1993 VW	&	0.930	$\pm$	0.001	&	2.000	$\pm$	0.002	&	1.07	$\pm$	0.01	\\
7148 Reinholdbien	&	0.928	$\pm$	0.008	&	1.942	$\pm$	0.021	&	1.01	$\pm$	0.03	\\
7800 Zhongkeyuan	&	0.921	$\pm$	0.002	&	1.932	$\pm$	0.019	&	1.01	$\pm$	0.02	\\
7889 1994 LX	&	0.930	$\pm$	0.003	&	1.937	$\pm$	0.005	&	1.01	$\pm$	0.01	\\
9481 Menchu	&	0.931	$\pm$	0.001	&	1.931	$\pm$	0.008	&	1.00	$\pm$	0.01	\\
9553 Colas	&	0.922	$\pm$	0.001	&	1.923	$\pm$	0.003	&	1.00	$\pm$	0.01	\\
10349 1992 LN	&	0.905	$\pm$	0.005	&	1.922	$\pm$	0.016	&	1.02	$\pm$	0.02	\\
11699 1998 FL105	&	0.916	$\pm$	0.002	&	1.938	$\pm$	0.003	&	1.02	$\pm$	0.01	\\
16416 1987 SM3	&	0.923	$\pm$	0.003	&	1.963	$\pm$	0.013	&	1.04	$\pm$	0.02	\\
16651 1993 TS11	&	0.925	$\pm$	0.005	&	1.910	$\pm$	0.011	&	0.99	$\pm$	0.02	\\
21238 Panarea	&	0.910	$\pm$	0.004	&	1.887	$\pm$	0.009	&	0.98	$\pm$	0.01	\\
26886 1994 TJ2	&	0.913	$\pm$	0.001	&	1.906	$\pm$	0.007	&	0.99	$\pm$	0.01	\\
27343 Deannashea	&	0.917	$\pm$	0.002	&	1.914	$\pm$	0.004	&	1.00	$\pm$	0.01	\\
33881 2000 JK66	&	0.929	$\pm$	0.001	&	1.930	$\pm$	0.004	&	1.00	$\pm$	0.01	\\
36412 2000 OP49	&	0.930	$\pm$	0.003	&	1.961	$\pm$	0.010	&	1.03	$\pm$	0.01	\\
38070 Redwine	&	0.932	$\pm$	0.002	&	1.958	$\pm$	0.007	&	1.03	$\pm$	0.01	\\
42947 1999 TB98	&	0.924	$\pm$	0.008	&	1.928	$\pm$	0.009	&	1.00	$\pm$	0.02	\\
50098 2000 AG98	&	0.925	$\pm$	0.002	&	1.947	$\pm$	0.004	&	1.02	$\pm$	0.01	\\
52750 1998 KK17	&	0.925	$\pm$	0.002	&	1.985	$\pm$	0.006	&	1.06	$\pm$	0.01	\\
66268 1999 JJ3	&	0.923	$\pm$	0.010	&	1.920	$\pm$	0.011	&	1.00	$\pm$	0.02	\\
88188 2000 XH44	&	0.920	$\pm$	0.002	&	1.975	$\pm$	0.003	&	1.06	$\pm$	0.01	\\
97276 1999 XC143	&	0.934	$\pm$	0.002	&	2.025	$\pm$	0.008	&	1.09	$\pm$	0.01	\\
137924 2000 BD19	&	0.937	$\pm$	0.005	&	1.952	$\pm$	0.010	&	1.02	$\pm$	0.01	\\
192563 1998 WZ6	&	0.920	$\pm$	0.003	&	2.005	$\pm$	0.014	&	1.09	$\pm$	0.02	\\
253841 2003 YG118	&	0.925	$\pm$	0.003	&	1.962	$\pm$	0.007	&	1.04	$\pm$	0.01	\\
297418 2000 SP43	&	0.938	$\pm$	0.002	&	2.038	$\pm$	0.012	&	1.10	$\pm$	0.01	\\
326290 Akhenaten	&	0.930	$\pm$	0.007	&	1.950	$\pm$	0.045	&	1.02	$\pm$	0.05	\\
2001 YE4	&	0.930	$\pm$	0.002	&	1.985	$\pm$	0.007	&	1.06	$\pm$	0.01	\\
2005 WX	&	0.930	$\pm$	0.013	&	1.933	$\pm$	0.028	&	1.00	$\pm$	0.04	\\
2008 BT18	&	0.919	$\pm$	0.001	&	1.955	$\pm$	0.006	&	1.04	$\pm$	0.01	\\
2011 YA	&	0.925	$\pm$	0.002	&	1.970	$\pm$	0.005	&	1.05	$\pm$	0.01	\\
2013 KL6	&	0.933	$\pm$	0.005	&	1.943	$\pm$	0.029	&	1.01	$\pm$	0.03	\\
\hline
\hline
Vestoids	&	0.920	 $\pm$	0.005	&	1.940	 $\pm$	0.015	&	1.02	 $\pm$	0.02	\\
Fugitives	&	0.925	 $\pm$	0.003	&	1.949	 $\pm$	0.013	&	1.02	 $\pm$	0.01	\\
IOs&			0.931	 $\pm$	0.006	&	1.954	 $\pm$	0.028	&	1.02	 $\pm$	0.02	\\
Low-i	&		0.925	 $\pm$	0.004	&	1.952	 $\pm$	0.019	&	1.03	 $\pm$	0.02	\\
NEAs	&		0.927	 $\pm$	0.004	&	1.968	 $\pm$	0.019	&	1.04	 $\pm$	0.02	\\
MOVs	&		0.919	 $\pm$	0.009	&	1.905	 $\pm$	0.018	&	0.99	 $\pm$	0.01	\\
\hline
\hline
Control sample	&	0.902 - 0.930		&	1.868 - 2.003	 &	0.97 - 1.10	\\
\end{longtable}
\end{center}
}

\begin{table*}
       \caption{Deconvolution of the V-type spectra according to the 1 or 2-pyroxene model of the MGM.  }
        \label{mgm1}
        \scriptsize{
\begin{tabular}{|l|l||c|c|c||c|c|c||c|c|c|} \hline
\hline
&	&		&Centres ($\mu m$)		&		&		&Widths ($\mu m$)		&		&		&Strengths 		&		\\
Group&	Object&	Extra M1	&	OPX-1	&	OPX-2	&	Extra M1	&	OPX-1	&	OPX-2	&	Extra M1	&	OPX-1	&	OPX-2	\\ \hline
Family&2045 Peking	&		&	0.921	&	1.947	&		&	0.247	&	0.564	&		&	-0.840	&	-0.465	\\
&2511 Patterson	&		&	0.920	&	1.939	&		&	0.241	&	0.599	&		&	-0.862	&	-0.533	\\
&3155 Lee	&		&	0.906	&	1.911	&		&	0.214	&	0.582	&		&	-0.823	&	-0.673	\\
&3657 Ermolova	&		&	0.915	&	1.912	&		&	0.242	&	0.495	&		&	-0.850	&	-0.610	\\
&3782 Celle	&		&	0.920	&	1.932	&		&	0.222	&	0.564	&		&	-0.700	&	-0.459	\\
&4215  Kamo	&		&	0.916	&	1.945	&		&	0.208	&	0.595	&		&	-0.588	&	-0.406	\\
&5111  Jacliff	&	0.799	&	0.933	&	1.930	&	0.249	&	0.181	&	0.398	&	-0.296	&	-0.493	&	-0.398	\\
&5481 Kiuchi	&		&	0.921	&	1.932	&		&	0.242	&	0.610	&		&	-0.969	&	-0.572	\\
\hline
Non-vestoids&2371 Dimitrov	&		&	0.923	&	1.956	&		&	0.226	&	0.590	&		&	-0.821	&	-0.492	\\
(low-i,&2851 Harbin	&		&	0.915	&	1.911	&		&	0.235	&	0.635	&		&	-0.980	&	-0.637	\\
IOs, MOVs,&2912 Lapalma	&		&	0.918	&	1.923	&		&	0.249	&	0.592	&		&	-1.008	&	-0.650	\\
fugitives)&7148   Reinholdbien	&		&	0.917	&	1.932	&		&	0.240	&	0.525	&		&	-0.911	&	-0.457	\\
&21238 Panarea	&		&	0.907	&	1.879	&		&	0.283	&	0.719	&		&	-1.035	&	-0.820	\\ \hline
NEAs&4055 Magellan	&		&	0.914	&	1.888	&		&	0.248	&	0.627	&		&	-0.995	&	-0.757	\\
&5604 1992FE	&		&	0.920	&	1.878	&		&	0.220	&	0.544	&		&	-0.795	&	-0.554	\\
\hline													
\hline	
\end{tabular}
} \raggedright
\smallskip
{\scriptsize \begin{tabular}{|l|l||c|c|c|c||c|c|c|c||c|c|c|c||c|}        
\hline\hline                 
& 	&	Cen. ($\mu m$)	&		&		&		&Wid. ($\mu m$)	&		&		&		&Stren.	&		&		&		&	CPX/	\\
Group&Object	&	OPX-1	&	CPX-1	&	OPX-2	&	CPX-2	&	OPX-1	&	CPX-1	&	OPX-2	&	CPX-2	&	OPX-1	&	CPX-1	&	OPX-2	&	CPX-2	&	(OPX+CPX)$^a$	\\ \hline
Family&1929 Kollaa	&	0.892	&	0.982	&	1.886	&	2.215	&	0.246	&	0.170	&	0.538	&	0.564	&	-0.787	&	-0.230	&	-0.508	&	-0.225	&	29-38	\\
&2011 Veteraniya	&	0.884	&	0.992	&	1.880	&	2.247	&	0.200	&	0.195	&	0.540	&	0.590	&	-0.677	&	-0.334	&	-0.371	&	-0.175	&	44-45	\\
\hline
Non-vestoids&809 Lundia	&	0.904	&	1.011	&	1.906	&	2.224	&	0.229	&	0.190	&	0.616	&	0.561	&	-0.965	&	-0.176	&	-0.624	&	-0.129	&	20-21	\\
(low-i,&854 Frostia	&	0.891	&	0.994	&	1.892	&	2.219	&	0.195	&	0.179	&	0.494	&	0.583	&	-0.369	&	-0.348	&	-0.344	&	-0.297	&	66-68	\\
IOs, MOVs,&956 Elisa	&	0.898	&	1.004	&	1.885	&	2.144	&	0.208	&	0.144	&	0.624	&	0.567	&	-0.894	&	-0.234	&	-0.546	&	-0.149	&	26-27	\\
fugitives)&1459 Magnya	&	0.894	&	0.996	&	1.838	&	2.042	&	0.192	&	0.118	&	0.565	&	0.412	&	-0.689	&	-0.333	&	-0.681	&	-0.342	&	45-47	\\
&2442 Corbett	&	0.882	&	0.988	&	1.853	&	2.183	&	0.204	&	0.160	&	0.541	&	0.547	&	-0.767	&	-0.342	&	-0.534	&	-0.273	&	42-47	\\
&2566 Kirghizia	&	0.898	&	0.979	&	1.928	&	2.462	&	0.228	&	0.172	&	0.606	&	0.628	&	-0.671	&	-0.195	&	-0.620	&	-0.195	&	28-31	\\
&2579 Spartacus	&	0.894	&	1.003	&	1.908	&	2.208	&	0.188	&	0.175	&	0.522	&	0.572	&	-0.648	&	-0.375	&	-0.419	&	-0.228	&	49-52	\\
&2653 Principia	&	0.907	&	0.975	&	1.900	&	2.270	&	0.224	&	0.194	&	0.559	&	0.586	&	-0.703	&	-0.161	&	-0.490	&	-0.199	&	23-38	\\
&2763 Jeans	&	0.904	&	0.970	&	1.915	&	2.237	&	0.220	&	0.205	&	0.537	&	0.581	&	-0.507	&	-0.277	&	-0.398	&	-0.199	&	46-49	\\
&2795 Lepage	&	0.898	&	0.988	&	1.873	&	2.202	&	0.231	&	0.181	&	0.564	&	0.563	&	-0.667	&	-0.175	&	-0.491	&	-0.221	&	28-38	\\
&38070 Redwine	&	0.891	&	0.988	&	1.852	&	2.150	&	0.200	&	0.172	&	0.542	&	0.548	&	-0.519	&	-0.300	&	-0.366	&	-0.231	&	52-55	\\ \hline
NEAs&1981 Midas	&	0.868	&	0.988	&	1.826	&	2.195	&	0.189	&	0.188	&	0.542	&	0.520	&	-0.558	&	-0.533	&	-0.430	&	-0.397	&	68	\\
&3908 Nyx		&	0.874	&	0.978	&	1.857	&	2.232	&	0.210	&	0.159	&	0.543	&	0.637	&	-0.785	&	-0.397	&	-0.633	&	-0.377	&	45-50	\\
&6611 1993VW	&	0.899	&	0.982	&	1.893	&	2.232	&	0.202	&	0.210	&	0.540	&	0.633	&	-0.473	&	-0.321	&	-0.357	&	-0.268	&	58-61	\\
&7889 1994LX	&	0.884	&	0.995	&	1.861	&	2.203	&	0.175	&	0.142	&	0.540	&	0.651	&	-0.811	&	-0.426	&	-0.634	&	-0.319	&	47-48	\\
\hline                                   
\end{tabular}
$^a$ These values represent the high calcium percentage pyroxene estimated trough the relative strengths of the OPX and CPX in both the 1 and 2 $\mu$m regions, using the method outlined by Sunshine \& Pieters (1993). Estimates are within $\pm$9\% of the actual values, according to data computed by Mayne et al. (2009) on eucrites.}
\end{table*}

\begin{table*}
\caption{VNIR spectral analysis: BI centre and depth, BII centre and depth, wollastonite [Wo] and ferrosilite [Fs] molar contents }.
\label{vnirtab}      
\centering
{
 \begin{tabular}{l c c c c c c}        
Object &	BI centre ($\mu m$) & 	BI depth&	BII centre ($\mu m$)&	BII depth&	[Wo](\%mol)&	[Fs](\%mol)\\
\hline\hline                 
809 Lundia	&	0.934	$\pm$	0.001	&	0.487	$\pm$	0.004	&	1.947	$\pm$	0.002	&	0.401	$\pm$	0.004	&	8.36	&	39.27	\\
854 Frostia	&	0.950	$\pm$	0.002	&	0.381	$\pm$	0.001	&	1.980	$\pm$	0.002	&	0.307	$\pm$	0.008	&	12.84	&	50.86	\\
956 Elisa 	&	0.930	$\pm$	0.001	&	0.476	$\pm$	0.006	&	1.933	$\pm$	0.002	&	0.406	$\pm$	0.004	&	7.00	&	35.78	\\
1459 Magnya	&	0.931	$\pm$	0.001	&	0.527	$\pm$	0.005	&	1.924	$\pm$	0.002	&	0.552	$\pm$	0.003	&	6.84	&	35.37	\\
1929 Kollaa	&	0.935	$\pm$	0.001	&	0.453	$\pm$	0.002	&	1.945	$\pm$	0.002	&	0.375	$\pm$	0.003	&	8.47	&	39.58	\\
1933 Tinchen 	&	0.924	$\pm$	0.002	&	0.322	$\pm$	0.009	&	1.915	$\pm$	0.007	&	0.192	$\pm$	0.002	&	5.10	&	30.86	\\
1981 Midas	&	0.935	$\pm$	0.002	&	0.460	$\pm$	0.003	&	1.990	$\pm$	0.010	&	0.335	$\pm$	0.006	&	10.27	&	44.21	\\
2011 Veteraniya	&	0.925	$\pm$	0.001	&	0.408	$\pm$	0.008	&	1.939	$\pm$	0.005	&	0.259	$\pm$	0.007	&	6.25	&	33.84	\\
2045 Peking 	&	0.933	$\pm$	0.001	&	0.424	$\pm$	0.008	&	1.946	$\pm$	0.003	&	0.333	$\pm$	0.009	&	8.12	&	38.66	\\
2371 Dimitrov	&	0.932	$\pm$	0.001	&	0.430	$\pm$	0.007	&	1.945	$\pm$	0.002	&	0.312	$\pm$	0.006	&	7.88	&	38.04	\\
2442 Corbett	&	0.930	$\pm$	0.002	&	0.473	$\pm$	0.007	&	1.942	$\pm$	0.002	&	0.389	$\pm$	0.002	&	7.36	&	36.71	\\
2511 Patterson	&	0.930	$\pm$	0.002	&	0.435	$\pm$	0.007	&	1.935	$\pm$	0.003	&	0.343	$\pm$	0.001	&	7.08	&	35.99	\\
2566 Kirghizia 	&	0.930	$\pm$	0.001	&	0.438	$\pm$	0.001	&	1.940	$\pm$	0.002	&	0.359	$\pm$	0.001	&	7.28	&	36.51	\\
2579 Spartacus	&	0.937	$\pm$	0.002	&	0.490	$\pm$	0.004	&	1.975	$\pm$	0.002	&	0.338	$\pm$	0.004	&	10.07	&	43.69	\\
2653 Principia 	&	0.930	$\pm$	0.002	&	0.464	$\pm$	0.003	&	1.955	$\pm$	0.003	&	0.364	$\pm$	0.006	&	7.88	&	38.05	\\
2763 Jeans	&	0.940	$\pm$	0.002	&	0.418	$\pm$	0.002	&	1.970	$\pm$	0.003	&	0.315	$\pm$	0.001	&	10.46	&	44.71	\\
2795 Lepage	&	0.935	$\pm$	0.002	&	0.428	$\pm$	0.006	&	1.940	$\pm$	0.002	&	0.367	$\pm$	0.002	&	8.27	&	39.06	\\
2851 Harbin	&	0.920	$\pm$	0.002	&	0.489	$\pm$	0.006	&	1.910	$\pm$	0.006	&	0.396	$\pm$	0.003	&	4.10	&	28.30	\\
2912 Lapalma	&	0.927	$\pm$	0.002	&	0.505	$\pm$	0.003	&	1.923	$\pm$	0.002	&	0.415	$\pm$	0.005	&	6.01	&	33.22	\\
3155 Lee	&	0.915	$\pm$	0.002	&	0.428	$\pm$	0.009	&	1.905	$\pm$	0.002	&	0.381	$\pm$	0.006	&	2.91	&	25.23	\\
3268 De Sanctis	&0.918 $\pm$		0.002	&0.298$\pm$		0.003	&1.981 $\pm$		0.005	&0.255 $\pm$		0.001&6.54   &34.58   \\
3498 Belton	&	0.922	$\pm$	0.002	&	0.280	$\pm$	0.008	&	1.905	$\pm$	0.005	&	0.319	$\pm$	0.002	&	4.30	&	28.81	\\
3657 Ermolova	&	0.932	$\pm$	0.001	&	0.455	$\pm$	0.007	&	1.921	$\pm$	0.002	&	0.433	$\pm$	0.001	&	6.92	&	35.57	\\
3782 Celle 	&	0.930	$\pm$	0.001	&	0.395	$\pm$	0.007	&	1.930	$\pm$	0.003	&	0.298	$\pm$	0.006	&	6.88	&	35.48	\\
3908 Nyx	&	0.930	$\pm$	0.002	&	0.557	$\pm$	0.008	&	1.920	$\pm$	0.010	&	0.453	$\pm$	0.005	&	6.48	&	34.45	\\
3968 Koptelov	&	0.928	$\pm$	0.001	&	0.349	$\pm$	0.005	&	1.918	$\pm$	0.004	&	0.227	$\pm$	0.005	&	6.01	&	33.22	\\
4038 Kristina	&0.925 $\pm$		0.001	&0.267 $\pm$		0.002	&1.968 $\pm$		0.003	&0.259 $\pm$		0.007&7.41 &36.83\\
4055 Magellan	&	0.920	$\pm$	0.002	&	0.481	$\pm$	0.005	&	1.920	$\pm$	0.010	&	0.441	$\pm$	0.009	&	4.50	&	29.33	\\
4215 Kamo	&	0.925	$\pm$	0.002	&	0.367	$\pm$	0.005	&	1.925	$\pm$	0.003	&	0.250	$\pm$	0.003	&	5.69	&	32.40	\\
5111 Jacliff	&	0.925	$\pm$	0.002	&	0.360	$\pm$	0.007	&	1.935	$\pm$	0.004	&	0.274	$\pm$	0.006	&	6.09	&	33.43	\\
5481 Kiuchi	&	0.931	$\pm$	0.001	&	0.461	$\pm$	0.010	&	1.931	$\pm$	0.002	&	0.370	$\pm$	0.010	&	7.12	&	36.09	\\
5604 1992 FE	&	0.925	$\pm$	0.002	&	0.476	$\pm$	0.005	&	1.910	$\pm$	0.010	&	0.347	$\pm$	0.007	&	5.09	&	30.86	\\
6611 1993 VW	&	0.935	$\pm$	0.002	&	0.446	$\pm$	0.006	&	1.975	$\pm$	0.010	&	0.296	$\pm$	0.003	&	9.67	&	42.67	\\
7148 Reinholbien	&	0.933	$\pm$	0.002	&	0.468	$\pm$	0.002	&	1.925	$\pm$	0.004	&	0.257	$\pm$	0.008	&	7.28	&	36.50	\\
7889 1994 LX	&	0.935	$\pm$	0.002	&	0.569	$\pm$	0.009	&	1.910	$\pm$	0.010	&	0.466	$\pm$	0.002	&	7.08	&	35.98	\\
11699 1998 FL105	&	0.928	$\pm$	0.002	&	0.391	$\pm$	0.008	&	1.938	$\pm$	0.002	&	0.354	$\pm$	0.006	&	6.81	&	35.28	\\
21238 Panarea	&	0.915	$\pm$	0.002	&	0.465	$\pm$	0.009	&	1.882	$\pm$	0.004	&	0.435	$\pm$	0.002	&	2.00	&	22.86	\\
38070 Redwine	&	0.936	$\pm$	0.002	&	0.407	$\pm$	0.004	&	1.946	$\pm$	0.003	&	0.328	$\pm$	0.006	&	8.71	&	40.19	\\
\hline
Vestoids	&	0.927	$\pm$	0.003	&	0.381	$\pm$	0.043	&	1.934	$\pm$	0.012	&	0.308	$\pm$	0.052	&6.36		&34.12		\\
Fugitives	&	0.933	$\pm$	0.003	&	0.476	$\pm$	0.011	&	1.941	$\pm$	0.011	&	0.364	$\pm$	0.027	&7.83		&37.92		\\
IOs	&	0.935	$\pm$	0.014	&	0.426	$\pm$	0.026	&	1.945	$\pm$	0.034	&	0.344	$\pm$	0.021	&	8.55	&39.78		\\
Low-i	&	0.932	$\pm$	0.004	&	0.445	$\pm$	0.020	&	1.950	$\pm$	0.005	&	0.348	$\pm$	0.030	&	8.17	&38.80		\\
NEAs	&	0.930	$\pm$	0.003	&	0.498	$\pm$	0.026	&	1.938	$\pm$	0.010	&	0.390	$\pm$	0.059	&	7.18	&36.25		\\
MOVs	&	0.923	$\pm$	0.008	&	0.496	$\pm$	0.031	&	1.903	$\pm$	0.021	&	0.494	$\pm$	0.059	&	4.42	&29.11		\\
\hline
V-types	&	0.929	$\pm$	0.005	&	0.432	$\pm$	0.035	&	1.937	$\pm$	0.013	&	0.347	$\pm$	0.048	&	7.03 $\pm$1.03	&35.85 $\pm$ 2.66\\
\hline                                   
\end{tabular}}
~\\
\raggedright
\smallskip
\end{table*}

{\scriptsize
       \begin{center}
     \begin{longtable} {|l|c|c|c|c|c|c|c|c|}  
\caption[]{VNIR spectral analysis: BI centre and depth, BII centre and depth, wollastonite [Wo] and ferrosilite [Fs] molar contents for a HED sample}. 
        \label{hd} \\
\hline 
\multicolumn{1}{|c|} {\textbf{Meteorite}} & \multicolumn{1}{c|}{\textbf{SubType}} & \multicolumn{1}{c|} {\textbf{Grainsize}} &\multicolumn{1}{c|} {\textbf{BI centre}} &     \multicolumn{1}{c|}{\textbf{BI depth}}  & \multicolumn{1}{c|} {\textbf{BII centre}} &     \multicolumn{1}{c|}{\textbf{BII depth}}  &\multicolumn{1}{c|} {\textbf{[Fs]}} &     \multicolumn{1}{c|}{\textbf{[Wo]}}\\  
\multicolumn{1}{|c|} {} & \multicolumn{1}{c|}{} & \multicolumn{1}{c|} {} &\multicolumn{1}{c|} {\textbf{($\mu m$)}} &     \multicolumn{1}{c|}{}  & \multicolumn{1}{c|} {\textbf{($\mu m$)}} &     \multicolumn{1}{c|}{}  & \multicolumn{1}{c|} {\textbf{(\%mol)}} &     \multicolumn{1}{c|}
{\textbf{(\%mol)}}\\
\hline 
\endfirsthead
\multicolumn{9}{c}%
{{\bfseries \tablename\ \thetable{} -- continued from previous page}} \\ \hline 
\endfoot
\hline 
\multicolumn{1}{|c|} {\textbf{Meteorite}} & \multicolumn{1}{c|}{\textbf{SubType}} & \multicolumn{1}{c|} {\textbf{Grainsize}} &\multicolumn{1}{c|} {\textbf{BI centre}} &     \multicolumn{1}{c|}{\textbf{BI depth}}  & \multicolumn{1}{c|} {\textbf{BII centre}} &     \multicolumn{1}{c|}{\textbf{BII depth}}  &\multicolumn{1}{c|} {\textbf{[Fs]}} &     \multicolumn{1}{c|}{\textbf{[Wo]}}\\  
\multicolumn{1}{|c|} {} & \multicolumn{1}{c|}{} & \multicolumn{1}{c|} {} &\multicolumn{1}{c|} {\textbf{($\mu m$)}} &     \multicolumn{1}{c|}{}  & \multicolumn{1}{c|} {\textbf{($\mu m$)}} &     \multicolumn{1}{c|}{}  & \multicolumn{1}{c|} {\textbf{(\%mol)}} &     \multicolumn{1}{c|}
{\textbf{(\%mol)}}\\
\hline   \hline
\endhead
\hline \multicolumn{9}{r}{{Continue on next page}} \\ 
\endfoot
\hline \hline
\endlastfoot
A-881526	&	D	&	$<25\mu$m	&	0.920	&	0.640	&	1.885	&	0.415	&	25.73	&	3.11	\\
Aioun el Atrouss 	&	D	&	$<25\mu$m	&	0.925	&	0.625	&	1.900	&	0.416	&	29.83	&	4.69	\\
EETA79002	&	D	&	$<25\mu$m	&	0.920	&	0.502	&	1.890	&	0.291	&	26.24	&	3.31	\\
GRO95555	&	D	&	$<25\mu$m	&	0.920	&	0.618	&	1.905	&	0.406	&	27.79	&	3.9	\\
Johnstown	&	D	&	$<25\mu$m	&	0.930	&	0.483	&	1.950	&	0.316	&	37.53	&	7.68	\\
LAP91900	&	D	&	$<25\mu$m	&	0.920	&	0.729	&	1.890	&	0.653	&	26.24	&	3.31	\\
Roda	&	D	&	$<25\mu$m	&	0.925	&	0.602	&	1.902	&	0.445	&	30.04	&	4.77	\\
Shalka 	&	D	&	$<25\mu$m	&	0.920	&	0.603	&	1.897	&	0.476	&	26.96	&	3.58	\\
Tatahouine	&	D	&	$<25\mu$m	&	0.920	&	0.625	&	1.895	&	0.406	&	26.76	&	3.5	\\
Y-74013	&	D	&	$<25\mu$m	&	0.925	&	0.414	&	1.920	&	0.242	&	31.89	&	5.49	\\
Y-75032	&	D	&	$<25\mu$m	&	0.925	&	0.511	&	1.935	&	0.299	&	33.43	&	6.09	\\ \hline
Binda	&	H	&	$<25\mu$m	&	0.925	&	0.650	&	1.935	&	0.444	&	33.43	&	6.09	\\
Bununu	&	H	&	$<25\mu$m	&	0.925	&	0.437	&	1.945	&	0.240	&	34.46	&	6.49	\\
EET83376	&	H	&	$<25\mu$m	&	0.935	&	0.453	&	1.965	&	0.289	&	41.64	&	9.27	\\
EET87503	&	H	&	$<25\mu$m	&	0.930	&	0.419	&	1.955	&	0.255	&	38.05	&	7.88	\\
EET87513	&	H	&	$<25\mu$m	&	0.935	&	0.364	&	1.965	&	0.188	&	41.64	&	9.27	\\
Frankfort	&	H	&	$<25\mu$m	&	0.930	&	0.509	&	1.940	&	0.313	&	36.51	&	7.28	\\
GRO95535	&	H	&	$<25\mu$m	&	0.930	&	0.420	&	1.950	&	0.243	&	37.53	&	7.68	\\
GRO95574	&	H	&	$<25\mu$m	&	0.930	&	0.478	&	1.940	&	0.313	&	36.51	&	7.28	\\
Kapoeta	&	H	&	$<25\mu$m	&	0.930	&	0.401	&	1.950	&	0.204	&	37.53	&	7.68	\\
Le Teilleul	&	H	&	$<25\mu$m	&	0.930	&	0.551	&	1.935	&	0.362	&	35.99	&	7.08	\\
Pavlovka	&	H	&	$<25\mu$m	&	0.920	&	0.527	&	1.927	&	0.393	&	30.05	&	4.78	\\
Petersburg	&	H	&	$<25\mu$m	&	0.935	&	0.432	&	1.970	&	0.275	&	42.15	&	9.47	\\
QUE94200	&	H	&	$<25\mu$m	&	0.925	&	0.475	&	1.920	&	0.296	&	31.89	&	5.49	\\
Y-7308	&	H	&	$<25\mu$m	&	0.925	&	0.571	&	1.935	&	0.368	&	33.43	&	6.09	\\
Y-790727	&	H	&	$<25\mu$m	&	0.930	&	0.484	&	1.950	&	0.316	&	37.53	&	7.68	\\
Y-791573	&	H	&	$<25\mu$m	&	0.925	&	0.462	&	1.940	&	0.287	&	33.95	&	6.29	\\ \hline
A-881819	&	E	&	$<25\mu$m	&	0.930	&	0.475	&	1.960	&	0.282	&	38.56	&	8.08	\\
ALH-78132	&	E	&	$<25\mu$m	&	0.930	&	0.421	&	1.960	&	0.262	&	38.56	&	8.08	\\
ALH-78132	&	E	&	 25-45$\mu m$	&	0.930	&	0.577	&	1.955	&	0.425	&		&		\\
ALH-78132	&	E	&	 45-75$\mu m$	&	0.930	&	0.614	&	1.960	&	0.496	&		&		\\
ALHA76005	&	E	&	$<25\mu$m	&	0.935	&	0.381	&	1.980	&	0.229	&	43.18	&	9.87	\\
ALHA76005	&	E	&	 25-45$\mu m$	&	0.935	&	0.489	&	1.980	&	0.364	&		&		\\
ALHA76005	&	E	&	 45-75$\mu m$	&	0.940	&	0.551	&	1.975	&	0.446	&		&		\\
ALHA81001	&	E	&	$<25\mu$m	&	0.935	&	0.311	&	2.005	&	0.218	&	45.75	&	10.87	\\
Bereba	&	E	&	$<25\mu$m	&	0.925	&	0.423	&	1.970	&	0.304	&	37.03	&	7.49	\\
Bouvante	&	E	&	$<45\mu$m	&	0.950	&	0.490	&	1.980	&	0.430	&	50.86	&	12.84	\\
EETA79005	&	E	&	$<25\mu$m	&	0.935	&	0.461	&	1.965	&	0.293	&	41.64	&	9.27	\\
Ibitira	&	E	&	$<25\mu$m	&	0.940	&	0.597	&	1.985	&	0.328	&	46.25	&	11.06	\\
Juvinas	&	E	&	$<25\mu$m	&	0.935	&	0.466	&	1.990	&	0.281	&	44.21	&	10.27	\\
Juvinas	&	E	&	 25-45$\mu m$	&	0.940	&	0.593	&	1.980	&	0.438	&		&		\\
Juvinas	&	E	&	 45-75$\mu m$	&	0.940	&	0.620	&	1.975	&	0.497	&		&		\\
LEW85303	&	E	&	$<25\mu$m	&	0.945	&	0.408	&	2.015	&	0.238	&	51.9	&	13.25	\\
LEW87004	&	E	&	$<25\mu$m	&	0.935	&	0.425	&	1.975	&	0.246	&	42.67	&	9.67	\\
Millbillillie	&	E	&	$<25\mu$m	&	0.940	&	0.336	&	2.005	&	0.206	&	48.31	&	11.86	\\
Millbillillie	&	E	&	 25-45$\mu m$	&	0.935	&	0.586	&	1.995	&	0.399	&		&		\\
Millbillillie	&	E	&	 45-75$\mu m$	&	0.940	&	0.591	&	1.995	&	0.434	&		&		\\
Nobleborough	&	E	&	$<25\mu$m	&	0.930	&	0.440	&	1.977	&	0.308	&	40.31	&	8.76	\\
Pasamonte	&	E	&	$<25\mu$m	&	0.935	&	0.402	&	1.992	&	0.264	&	44.42	&	10.35	\\
PCA82501	&	E	&	$<25\mu$m	&	0.945	&	0.337	&	1.985	&	0.229	&	48.81	&	12.05	\\
PCA82502	&	E	&	$<25\mu$m	&	0.940	&	0.462	&	2.005	&	0.314	&	48.31	&	11.86	\\
PCA91007	&	E	&	$<25\mu$m	&	0.950	&	0.458	&	1.990	&	0.404	&	51.89	&	13.24	\\
Serra de Mage	&	E	&	$<25\mu$m	&	0.930	&	0.439	&	1.960	&	0.270	&	38.56	&	8.08	\\
Sioux County	&	E	&	$<25\mu$m	&	0.930	&	0.356	&	1.970	&	0.279	&	39.59	&	8.48	\\
Stannern	&	E	&	$<25\mu$m	&	0.940	&	0.402	&	2.000	&	0.244	&	47.8	&	11.66	\\
Stannern	&	E	&	 25-45$\mu m$	&	0.940	&	0.518	&	1.995	&	0.398	&		&		\\
Y-74450	&	E	&	$<25\mu$m	&	0.930	&	0.376	&	1.980	&	0.193	&	40.62	&	8.88	\\
Y-74450	&	E	&	 25-45$\mu m$	&	0.935	&	0.530	&	1.970	&	0.342	&		&		\\
Y-74450	&	E	&	 45-75$\mu m$	&	0.935	&	0.568	&	1.970	&	0.407	&		&		\\
Y-792769	&	E	&	$<25\mu$m	&	0.940	&	0.397	&	2.005	&	0.217	&	48.31	&	11.86	\\
Y-793591	&	E	&	$<25\mu$m	&	0.940	&	0.399	&	1.995	&	0.241	&	47.28	&	11.46	\\
Y-82082	&	E	&	$<25\mu$m	&	0.945	&	0.384	&	2.010	&	0.228	&	51.39	&	13.05	\\

\end{longtable}
\end{center}}


\bsp	
\label{lastpage}
\end{document}